\RequirePackage{rotating}
\documentclass[acmsmall, screen, 10pt]{acmart}
\setcopyright{rightsretained}
\acmPrice{}
\acmDOI{10.1145/3428198}
\acmYear{2020}
\copyrightyear{2020}
\acmSubmissionID{oopsla20main-p19-p}
\acmJournal{PACMPL}
\acmVolume{4}
\acmNumber{OOPSLA}
\acmArticle{130}
\acmMonth{11}

\usepackage{listings}
\usepackage{dsfont}
\lstdefinelanguage{qsharp}
{
  morekeywords={
    operation,
    body,
    let,
    for,
    if, 
    open, 
    namespace, 
    function
  },
  sensitive=false, 
  morecomment=[l]{//}, 
  morecomment=[s]{/*}{*/}, 
  morestring=[b]" 
}
\usepackage{color}
\definecolor{eclipseBlue}{RGB}{42,0.0,255}
\definecolor{eclipseGreen}{RGB}{63,127,95}
 
\lstdefinestyle{qsharpstyle}{
  language={qsharp},
  basicstyle=\footnotesize\ttfamily, 
  captionpos=b, 
  extendedchars=true, 
  tabsize=2, 
  columns=fixed, 
  keepspaces=true, 
  showstringspaces=false, 
  breaklines=true, 
  framesep=4pt, 
  numbers=left, 
  numberstyle=\tiny\textrm, 
  commentstyle=\color{eclipseGreen}, 
  keywordstyle=\bfseries, 
  backgroundcolor =\color{white},
  numbersep=5pt,
  xleftmargin=3ex
}
\lstdefinestyle{cppstyle}{
	language={c++},
	basicstyle=\footnotesize\ttfamily, 
	captionpos=b, 
	extendedchars=true, 
	tabsize=2, 
	columns=fixed, 
	keepspaces=true, 
	showstringspaces=false, 
	breaklines=true, 
	framesep=4pt, 
	numbers=left, 
	numberstyle=\tiny\textrm, 
	commentstyle=\color{eclipseGreen}, 
	keywordstyle=\bfseries, 
	backgroundcolor =\color{white},
	numbersep=5pt,
	xleftmargin=3ex
}

\usepackage{booktabs}   
\usepackage{braket}
\usepackage{enumerate}
\usepackage{mathtools}
\usepackage{listings}
\usepackage{xcolor}
\usepackage{tikz}
\usepackage{pgfplots}
\usepackage{graphicx}
\usepackage[caption=false]{subfig}
\usetikzlibrary{positioning}
\usepackage{bbm}
\usepackage{amsmath}
\usepackage{algorithm}
\usepackage[noend]{algpseudocode}
\usepackage{float}
\usepackage{multicol}
\usepackage{rotating}
\usepackage{afterpage}
\usepackage{capt-of}
\usepackage{microtype}

\usepgfplotslibrary{groupplots}
\pgfplotsset{
  compat=newest,
  xlabel near ticks,
  ylabel near ticks
}
\newtheorem{example}{Example}

\begin{document}

\title[Enabling Accuracy-Aware Quantum Compilers using Symbolic Resource Estimation]{Enabling Accuracy-Aware Quantum Compilers\\using Symbolic Resource Estimation}    
   
\author{Giulia Meuli}
\affiliation{ EPFL
	\city{Lausanne}
	\country{Switzerland}
}
\email{giulia.meuli@epfl.ch}         

\author{Mathias Soeken}
\affiliation{ Microsoft
	\city{Zurich}
	\country{Switzerland}                   
}
\email{masoeken@microsoft.com}

\author{Martin Roetteler}
\affiliation{ Microsoft
	\city{Redmond, WA}
	\country{United States of America}
}
\email{martinro@microsoft.com}

\author{Thomas H\"aner}
\affiliation{ Microsoft
	\city{Zurich}
	\country{Switzerland}
}
\email{thhaner@microsoft.com}

\begin{abstract}
Approximation errors must be taken into account when compiling quantum programs into a low-level gate set. We present a methodology that tracks such errors automatically and then optimizes accuracy parameters to guarantee a specified overall accuracy while aiming to minimize the implementation cost in terms of quantum gates.
The core idea of our approach is to extract functions that specify the optimization problem directly from the high-level description of the quantum program. Then, custom compiler passes optimize these functions, turning them into (near-)symbolic expressions for (1) the total error and (2) the implementation cost (e.g., total quantum gate count). All unspecified parameters of the quantum program will show up as variables in these expressions, including accuracy parameters.
After solving the corresponding optimization problem, a circuit can be instantiated from the found solution.
We develop two prototype implementations, one in C++ based on Clang/LLVM, and another using the Q\# compiler infrastructure. We benchmark our prototypes on typical quantum computing programs, including the quantum Fourier transform, quantum phase estimation, and Shor's algorithm.
\end{abstract}

\ccsdesc[500]{Software and its engineering~Compilers}
\ccsdesc[500]{Hardware~Quantum computation}

\keywords{quantum computing, quantum programming, quantum algorithms, approximation errors, resource estimation}  

\maketitle

\section{Introduction}\label{sec:intro}
There exists a wide range of quantum algorithms that promise asymptotic speed-ups with respect to their classical counterparts. Application domains include crypto\-graphy~\cite{pirandola19, bennett14}, machine learning~\cite{Havl19}, material science~\cite{mcardle20}, and quantum chemistry~\cite{low19, babbush17}. However, concrete resource estimates for problems and corresponding problem sizes at which quantum computers are expected to outperform their classical counterparts remain scarce.
To carry out such resource estimates, several quantum programming languages and toolchains have been developed such as Q\#~\cite{SGT+18}, Quipper~\cite{quipper}, Scaffold/ScaffCC~\cite{scaffCC}, Qiskit~\cite{qiskit}, ProjectQ~\cite{projectQ}, and QuRE~\cite{SKF+13}. Some of these frameworks provide domain-specific languages with the necessary abstractions, libraries, simulators, and cloud access to small-scale quantum computers.
Despite the availability of these languages, there is still a significant amount of manual work involved in resource estimation~\cite{reiher17, Scherer17}---one reason being the lack of built-in support for handling approximation errors. 

Why do approximation errors occur in quantum programs in the first place? We discuss three main sources of errors in this paper, however, we note that the framework is extensible to other sources of errors.

\paragraph{1.~Synthesis Errors.} Due to the discrete nature of fault-tolerant instruction sets (and indeed, it is known that any universal fault-tolerant instruction set necessarily must be discrete~\cite{NC2000}), it cannot be avoided to introduce approximation errors for general target operations. For instance, consider a rotation around the $Z$-axis such as 
\[ R_Z(\theta) = \left[ \begin{array}{cc} e^{-i \theta/2} & 0 \\ 0 & e^{i \theta/2} \end{array} \right],
\] 
which can be defined for any $\theta \in [0, 4\pi)$. These rotations can only be implemented exactly for a discrete subset of the interval $[0,4\pi)$, as gates have to be expressed as words of finite length over any universal set of generators. It should be noted that there is a mathematical function that expresses the length of the approximating word in terms of an approximation error, which we call $\varepsilon_R$ in this paper. This mathematical function depends on the concrete synthesis algorithm used to perform the factorization into fault-tolerant instructions. State-of-the-art synthesis algorithms lead to a cost (e.g., number of $T$ gates, where $T = e^{i\pi/8}R_Z(\pi/4)$) that is proportional to $\log_2(\varepsilon_R^{-1})$ \cite{KMM2013,RS2016}.

\paragraph{2.~Phase Estimation Errors.} An important technique in quantum computing is to extract estimates of an eigenvalue $\lambda$ of an operator $U$~\cite{Shor94,Kitaev95} to $k$ bits of precision. A common method to achieve this is to prepare an eigenstate $\ket{\psi_\lambda}$ of $U$ and to then apply powers $U^{2^i}$, for $i=0, \ldots, k{-}1$, to the eigenstate $\ket{\psi_\lambda}$. This application is done conditionally on the value of a reference system and allows us to extract the $k$ most significant bits of the eigenvalue. As $\lambda$ can in principle be any complex number of the form $\lambda = e^{i \alpha}$, where $\alpha \in [0, 2\pi)$, the particular choice of $k$ introduces an approximation error and limits how precisely we can estimate $\lambda$. We call the resulting approximation error $\varepsilon_{QPE}$ in this paper. 

\paragraph{3.~Algorithmic Errors.} Some quantum programs are part of a parametric family of programs that gracefully degenerate with a reduction of the parameters. A concrete example for such a family of programs is the quantum Fourier transform~\cite{Shor94}, or QFT for short. While the transformation itself can be implemented exactly and with no approximation error over a gate set that includes continuous rotations such as $R_Z(\theta)$ for arbitrary $\theta\in [0,4\pi)$, it is possible to approximate the transformation by selectively dropping some of the rotations that occur, in particular by ``pruning'' the values of $\theta$ that are very close to $0$. One such pruning method is well known~\cite{coppersmith02} and allows us to drop many of the $O(n^2)$ rotations that a simple implementation of the Fourier transform requires and just retain $O(n\log{n})$ rotations, while still maintaining a sufficient approximation. We call the resulting approximation error $\varepsilon_{QFT}$ in this paper. Another example of algorithmic error comes from formulas that are known to converge to a target program when taking a suitable limit, e.g., of alternations of other, typically smaller and simpler, programs. An example for the latter is the so-called Trotter formula, a well-known identity to implement an approximation to $e^{i (A+B)}$ for Hermitian matrices $A$ and $B$, from the knowledge of implementations for $e^{iA}$ and $e^{iB}$. The resulting approximation error (the ``Trotter error'') is called $\varepsilon_{TE}$ in this paper. 

\subsection{Overview of Our Approach} A quantum program can be expressed at a high level of abstraction using any of the languages and frameworks mentioned above. Instead of specifying quantum circuits at the level of single- and two-qubit quantum gates, these languages provide high-level abstractions such as \textit{quantum Fourier transform} and \textit{quantum phase estimation}, which can be used to express quantum algorithms. Once the target machine has been specified, a compiler is used to translate these high-level abstractions into lower-level gates, such as the gates from the Clifford+$T$ gate set~\cite{Amy13}, which can be implemented fault-tolerantly on several scalable quantum computer architectures~\cite{BK05}. 

Various approximations, such as the ones described above, may be necessary during compilation and 
with existing languages, programmers must manually keep track of all the introduced errors.
Furthermore, programmers must tune the parameters of their implementation to keep the total error within a target budget. To guarantee this, they must derive the resulting error bounds manually---a task that is tedious and error prone. 

To address this issue, our methodology introduces language support into existing quantum programming languages, allowing programmers to deal independently with the approximation errors in each subroutine. The job of inferring how all introduced approximation errors interact is thus transferred from the programmer to the compiler. Our methodology automatically infers an error bound for the overall quantum program and then selects appropriate values for each of the program's accuracy parameters to simultaneously (1) satisfy a user-specified overall tolerance and (2) reduce the required quantum resources.

More specifically, our methodology supports:
\begin{enumerate}
\item {given the desired approximation error, determining the assignment of accuracy parameters that guarantees the given approximation error while aiming to minimize the number of operations;}
\item {given a maximal operation count, determining the assignment of accuracy parameters that yield at most the given operation count while aiming to reduce the total approximation error.}
\end{enumerate}

The automatic optimization of accuracy parameters is carried out by solving an optimization problem before the quantum circuit is generated. The constraint and cost functions describing the optimization problem are extracted by the compiler directly from the source code of the quantum program. Then, before execution, the optimized accuracy parameters are fed into the main program.

Finding a suitable assignment of accuracy parameters using, e.g., simulated annealing, requires hundreds of evaluations of both constraint and cost functions. Hence, our methodology has to lean on a fast method to estimate resource requirements and the total approximation error.
Available methods, e.g., in Q\#, estimate resources by actually generating quantum circuits from completely specified programs and then counting the generated gates. Hence, their runtime will increase with increasing problem size, making them ill-suited for use as cost functions in an optimization procedure, as we will show in Section~\ref{sect:runtime}.

Instead, we propose fast symbolic methods that ex\-tract a symbolic expression for the desired cost or constraint function (total approximation error or number of gates) directly from the source code of the quantum program. The resulting expressions feature variables that correspond to the various parameters of the program, including accuracy parameters. The symbolic approach does not need to execute the complete control flow of the quantum program to get an estimate, hence it provides a much faster solution that is viable even for application-scale programs. 

Our two prototypes both implement a symbolic approach for resource and error estimation. Since the resulting expressions may theoretically still contain some residual code that must be executed (e.g., certain if-else statements), we refer to them as being \textit{(near-)symbolic}. 
The parentheses indicate that, for most applications, the resulting expressions would be fully symbolic. However, there are examples where this is not the case, e.g., a quantum program that executes one of two different algorithms depending on a runtime parameter.
We note that our prototypes both generate fully symbolic expressions for all the examples in this paper.

\subsection{Contributions} To the best of our knowledge, we are the first to present a quantum programming framework that provides built-in support for automatic accuracy management. 
Our methodology automatically selects accuracy parameters such that the overall error is at most equal to a user-specified value while aiming to reduce the quantum resource requirements, or vice versa. As the interplay between the various approximation errors can be quite complicated, it can be difficult for a human to find the best trade-offs.

In addition, our approach for extracting \emph{symbolic} resource estimates from the quantum program, and using them to specify and solve the optimization problem of tuning accuracy parameters, appears to be new. 
Some state-of-the-art methods exist to automate resource estimation. Examples are methods embedded in Q\#, ProjectQ, and Quipper~\cite{SRSV14}, which are based on circuit-description languages, and QuRE, which is capable of evaluating different technologies and error-correcting codes. 
Nevertheless, all these frameworks require \emph{a priori} specified quantum circuits.
Thus, they cannot be used to derive asymptotic estimates as a function of the input parameters of the algorithm, in contrast to the symbolic approach we propose.

Indeed, a salient feature is that we acquire symbolic estimates for the total error and gate count directly from the high-level description of a quantum program, without executing the control-flow. It is only after the final resource requirements and total approximation error are known that we instantiate a circuit using the determined accuracy parameters. This allows us to automatically evaluate the resource requirements of a given quantum program in an accuracy-aware fashion, without the overhead of having to execute the entire control-flow of the program.

We implement a prototype of our framework in C++ for evaluation purposes. To demonstrate integration into an existing quantum programming language, we then present a prototype implementation into the Q\# compiler. In both cases we implement new compiler passes to enable the new functionality. Finally, we validate and benchmark our methodology on quantum programs such as \textit{quantum phase estimation} and \textit{Shor's algorithm}~\cite{Shor94}. We show that the runtime gap between the best previous methods and our symbolic method for resource and error estimation is {\em unbounded} as a function of the problem size.

Our methodology can be integrated into any of the cited quantum programming frameworks. We give details on the language and compiler features that are required to achieve such integration. Furthermore, our methodology is fully extensible: (1) more algorithmic errors can be defined at any point in the quantum program and will be handled automatically by our implementation; (2) more gate errors can be defined by adding a new type of error to each primitive gate; (3) custom implementations of library functions can be added.

\subsection{Related Work}
In the work by~\citet{HHZ+19}, a theoretical framework is presented to reason about the robustness of quantum programs when executed on noisy quantum hardware. Specifically, the authors develop a logic with which it is possible to characterize the distance between an ideal quantum program and its noisy counterpart, given a noise model of the target hardware. \citet{HHZ+19} present several case studies consisting of small quantum circuits (between 1 and 6 qubits). Computing the $(Q, \lambda)$-diamond norm, which measures the distance between the ideal and the erroneous program, involves solving a semidefinite programming problem (as is the case for the regular diamond norm~\cite{watrous2009semidefinite}). This becomes computationally intractable for large systems (and their corresponding noise models) due to the exponential scaling of the dimension with the number of qubits.

In contrast, our framework is concerned with approximation errors that occur at the algorithmic level. In particular, such errors may be decreased by using more quantum resources. Our framework thus aims to select appropriate accuracy parameters balancing this trade-off. Furthermore, we present two prototype implementations of our framework and demonstrate that it is capable of handling application-scale quantum programs featuring thousands of qubits and billions of operations.

Our methodology also relates to the automatic approach presented by~\citet{panchekha15} for handling floating-point rounding errors in classical computing. Their approach, called \emph{Herbie}, is capable of locating sources of errors in the code and proposing candidate rewrites to improve the overall precision. Similarly, our approach locates and adapts accuracy parameters to reduce the total error. While Herbie must be combined with other methods~\cite{darulova14, boldo08, barr13} to provide worst-case guarantees, our methodology proposes parameter assignments that guarantee the specified worst-case upper bound, assuming a correct implementation. We are not concerned with verifying actual correctness of the quantum program, but refer the reader to, e.g., the works by~\citet{Ying2011,PRZ2017,ARS2017}.

Symbolic execution, which is widely used for code testing, is another approach related to our work. The method translates the given program into a logical formula in order to check some input properties. A relevant example is veritesting~\cite{ARCB14}, which alternates dynamic (DSE) and static (SSE) symbolic execution. Similarly, we extract from a quantum program symbolic expressions for upper bounds on the total error and the gate count. 

\subsection{Paper Outline}
This paper is organized as follows. 

\noindent Section~\ref{sec:preliminaries} provides the necessary background information on quantum computing. In particular, Section~\ref{sect:errorqc} provides a detailed description of how approximation errors compose in quantum programs.

\noindent Section~\ref{sec:sample} illustrates three typical quantum algorithms, with an emphasis on the trade-offs between implementation cost and approximation error.

\noindent Section~\ref{sect:support} motivates the need for language support when implementing large-scale quantum algorithms in an accuracy-aware fashion. Furthermore, it illustrates how the proposed language features improve the readability and simplicity of the code, thus simplifying the implementation process.

\noindent Section~\ref{sec:automating} describes the proposed procedure to extract expressions for cost and constraint functions from a given quantum program. These expressions define an optimization problem, the solution of which yields appropriate values for all accuracy parameters. Moreover, the section details how the extracted expressions are optimized in order to arrive at (near-)symbolic expressions that may be evaluated much faster.

\noindent Section~\ref{sec:comp} provides a list of all the features that a quantum programming language and its compiler should support to integrate our proposed methodology. It also contains a description of how these features have been implemented in our LLVM prototype and in our Q\#-based implementation.

\noindent Section~\ref{sect:qualitative} shows the pseudo-code of our sample quantum programs and the corresponding  cost and constraint functions that our framework automatically deduces from the code.

\noindent Section~\ref{sect:runtime} reports the results of our runtime evaluation, which compares solving the optimization problem using a simulated annealing procedure with or without extracting (near-)symbolic expressions. It shows that our approach of extracting (near-)symbolic expressions is capable of reducing the time to solution by many orders of magnitude.

\section{Quantum Computing Background}
\label{sec:preliminaries}

\subsection{Quantum States}
Quantum computers process information encoded in qubits. The quantum state of a qubit $\ket{\psi}$ (using \textit{Dirac} or \textit{bra-ket} notation) can be written as
\[
	\ket\psi = \alpha_0\ket0 + \alpha_1\ket1,
\]
where $\alpha_0,\alpha_1\in \mathds{C}$ are the complex probability amplitudes corresponding to the \textit{computational basis states} $\ket0$ and $\ket1$, respectively, and $|\alpha_0|^2+|\alpha_1|^2=1$.
The computational basis states $\ket0$ and $\ket1$ may be associated with two-dimensional basis vectors $|0\rangle = \begin{psmallmatrix} 1 \\ 0 \end{psmallmatrix}$ and $|1\rangle = \begin{psmallmatrix} 0 \\ 1 \end{psmallmatrix}$. Single-qubit quantum gates (or quantum operations) can be written as $2\times 2$-dimensional complex unitary matrices. A matrix $U$ is unitary if its conjugate transpose corresponds to its inverse: $UU^{\dagger} = U^{\dagger}U = \mathds{1}$.
When applying a quantum gate to a single qubit, its new quantum state can be computed via a matrix-vector multiplication of the gate matrix and the two-dimensional amplitude vector $|\psi\rangle = \begin{psmallmatrix} \alpha_0 \\ \alpha_1 \end{psmallmatrix}$.

Composition of vector spaces is achieved using the \emph{tensor product}. Suppose $V$ and $W$ are Hilbert spaces of dimension $m$ and $n$ respectively. Then $V\otimes W$ is an $n\times m$-dimensional vector space. To see the connection to quantum states, let $V$ denote a two-dimensional vector space with basis vectors $\ket{0}$ and $\ket{1}$ (a single-qubit state), then $\ket{0}\otimes\ket{0} + \ket{1}\otimes\ket{1} $ is an element of $V\otimes V$ (i.e., a 2-qubit state).
The notation $\ket\psi^{\otimes k}$ means $\ket\psi$ tensored with itself $k$ times. An analogous notation is also used for operators on tensor product spaces. 
A more detailed description of the tensor product and its properties can be found in~\citet[Section 2.1.7]{NC2000}.

Therefore, an $n$-qubit quantum state may be described by a vector containing the $2^n$ amplitudes corresponding to all possible bitstrings of length $n$, or, using the bra-ket notation and two shorter alternatives:
\[
\sum_{x\in \mathds{B}^n} a_x |x_1\rangle\otimes \cdots \otimes|x_n\rangle = \sum_{x\in \mathds{B}^n} a_x |x_1\cdots x_n\rangle = \sum_{x\in \mathds{B}^n} a_x |x\rangle,
\]
where $x_i$ denotes the $i$-th bit of $x$ and $\sum_x|a_x|^2=1$. Measuring all $n$ qubits of the $n$-qubit state above results in a \textit{collapse} of the superposition onto one of the computational basis states $\ket x$ with probability $|a_x|^2$.

\subsection{Quantum Programs and Gates}
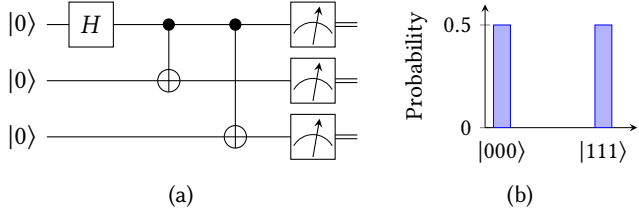
\begin{figure}
\centering
\subfloat[] {\begin{tikzpicture}[scale=1.400000,x=1pt,y=1pt]
\filldraw[color=white] (0.000000, -7.500000) rectangle (84.000000, 37.500000);
\draw[color=black] (0.000000,30.000000) -- (72.000000,30.000000);
\draw[color=black] (72.000000,29.500000) -- (84.000000,29.500000);
\draw[color=black] (72.000000,30.500000) -- (84.000000,30.500000);
\draw[color=black] (0.000000,30.000000) node[left] {$|0\rangle$};
\draw[color=black] (0.000000,15.000000) -- (72.000000,15.000000);
\draw[color=black] (72.000000,14.500000) -- (84.000000,14.500000);
\draw[color=black] (72.000000,15.500000) -- (84.000000,15.500000);
\draw[color=black] (0.000000,15.000000) node[left] {$|0\rangle$};
\draw[color=black] (0.000000,0.000000) -- (72.000000,0.000000);
\draw[color=black] (72.000000,-0.500000) -- (84.000000,-0.500000);
\draw[color=black] (72.000000,0.500000) -- (84.000000,0.500000);
\draw[color=black] (0.000000,0.000000) node[left] {$|0\rangle$};
\begin{scope}
\draw[fill=white] (12.000000, 30.000000) +(-45.000000:8.485281pt and 8.485281pt) -- +(45.000000:8.485281pt and 8.485281pt) -- +(135.000000:8.485281pt and 8.485281pt) -- +(225.000000:8.485281pt and 8.485281pt) -- cycle;
\clip (12.000000, 30.000000) +(-45.000000:8.485281pt and 8.485281pt) -- +(45.000000:8.485281pt and 8.485281pt) -- +(135.000000:8.485281pt and 8.485281pt) -- +(225.000000:8.485281pt and 8.485281pt) -- cycle;
\draw (12.000000, 30.000000) node {$H$};
\end{scope}
\draw (33.000000,30.000000) -- (33.000000,15.000000);
\begin{scope}
\draw[fill=white] (33.000000, 15.000000) circle(3.000000pt);
\clip (33.000000, 15.000000) circle(3.000000pt);
\draw (30.000000, 15.000000) -- (36.000000, 15.000000);
\draw (33.000000, 12.000000) -- (33.000000, 18.000000);
\end{scope}
\filldraw (33.000000, 30.000000) circle(1.500000pt);
\draw (51.000000,30.000000) -- (51.000000,0.000000);
\begin{scope}
\draw[fill=white] (51.000000, 0.000000) circle(3.000000pt);
\clip (51.000000, 0.000000) circle(3.000000pt);
\draw (48.000000, 0.000000) -- (54.000000, 0.000000);
\draw (51.000000, -3.000000) -- (51.000000, 3.000000);
\end{scope}
\filldraw (51.000000, 30.000000) circle(1.500000pt);
\draw[fill=white] (66.000000, 24.000000) rectangle (78.000000, 36.000000);
\draw[very thin] (72.000000, 30.600000) arc (90:150:6.000000pt);
\draw[very thin] (72.000000, 30.600000) arc (90:30:6.000000pt);
\draw[->,>=stealth] (72.000000, 24.600000) -- +(80:10.392305pt);
\draw[fill=white] (66.000000, 9.000000) rectangle (78.000000, 21.000000);
\draw[very thin] (72.000000, 15.600000) arc (90:150:6.000000pt);
\draw[very thin] (72.000000, 15.600000) arc (90:30:6.000000pt);
\draw[->,>=stealth] (72.000000, 9.600000) -- +(80:10.392305pt);
\draw[fill=white] (66.000000, -6.000000) rectangle (78.000000, 6.000000);
\draw[very thin] (72.000000, 0.600000) arc (90:150:6.000000pt);
\draw[very thin] (72.000000, 0.600000) arc (90:30:6.000000pt);
\draw[->,>=stealth] (72.000000, -5.400000) -- +(80:10.392305pt);
\draw[white] (0, -9)-- (2, -9);
\end{tikzpicture}}
\hspace{10pt}
\subfloat [] {\scalebox{.9}{\begin{tikzpicture}
\begin{axis} [
ybar ,
ylabel={Probability},
xtick=data,
xticklabels={$\ket{000}$,$\ket{111}$},
xlabel near ticks,
xtick style={draw=none},
ymin=0,
ymax=0.6,
xmin=-0.5,
xmax=4,
ytick={0,.5},
bar width = 0.5,
x=0.5cm,
y=3cm,
axis x line=bottom,
axis y line=left
]
\addplot coordinates { (0, 0.5) (3, 0.5) };
\end{axis}
\end{tikzpicture}}}
\caption{Example of a quantum circuit computing an entangled state (a) and the corresponding expected measurement outcomes (b).}
\label{exqc}
\end{figure}

A quantum program consists of classical \textit{and} quantum instructions~\cite{knill96}. While classical instructions are performed on the (classical) controller, the latter are performed on the quantum processing unit (QPU). In each step of the computation, the controller sends a sequence of quantum instructions to the QPU. After executing each sequence, the QPU returns classical measurement results to the controller, which may use them to decide on the quantum instructions to perform next (if any).

The mentioned sequences of quantum instructions can be visualized using circuit diagrams, so-called \textit{quantum circuits}.
\begin{example}
An example of a quantum circuit is shown in Fig.~\ref{exqc}. This circuit generates a Greenberger\---Horne\---Zeilinger state on three qubits, i.e., the state
\[
	\frac1{\sqrt 2}(\ket{000}+\ket{111}).
\]
In a quantum circuit, each qubit is represented by a horizontal line. Operations are denoted by boxes or other symbols on the qubit(s) they are being applied to. Time advances from left to right. The initial state is $\ket{0}^{\otimes 3}$.
The first operation is a Hadamard gate ($H$) applied to the first qubit. In matrix notation,
\[
H = \frac1{\sqrt{2}} \begin{bmatrix}
1 & 1 \\ 1 & -1
\end{bmatrix}.
\]
Applying $H$ to the first qubit maps $\ket{000}$ to $\frac1{\sqrt{2}}(|000\rangle + |100\rangle)$.
The next gate is a controlled-NOT or $\mathrm{CNOT}$,
\[
\mathrm{CNOT} = \begin{bsmallmatrix} 1 & 0 & 0 & 0 \\ 0 & 1 & 0 & 0 \\ 0 & 0 & 0 & 1 \\ 0 & 0 & 1 & 0 \end{bsmallmatrix},
\]
which entangles the first with the second qubit by flipping the latter if the first qubit is $\ket 1$. After the two $\mathrm{CNOT}$ gates, the three qubits are in the state $\frac1{\sqrt{2}} (|000\rangle + |111\rangle)$. Finally, all three qubits are measured. There is a 50\% probability of measuring all 0s and a 50\% probability of measuring all 1s. 
\end{example}

Quantum programs can be written in one of the various languages and software frameworks for quantum computing, e.g., ProjectQ, Q\#, Qiskit.
All of these languages allow programmers to specify the quantum program in terms of high-level operations. They also provide methods to decompose such operations into the native gate set of the QPU. 

Different quantum architectures support different native operations. A common low-level instruction set is the so-called Clifford+$T$ gate set~\cite{Amy13}. 
This instruction set includes the Clifford gates, i.e., the $\mathrm{CNOT}$ gate and Hadamard gate $H$, as well as the phase gate $S$, plus the $T$ gate, which is a non-Clifford gate required to achieve universality:

\[
S = \begin{bmatrix}
1 & 0 \\ 0 & i
\end{bmatrix}, \,
T = \begin{bmatrix}
1 & 0 \\ 0 & e^{i \pi/4} \end{bmatrix}. 
\]

In a fault-tolerant setting, it is particularly expensive to apply the $T$ gate. Consequently, the $T$-count (number of $T$ gates) is a good measure for the cost of a fault-tolerant implementation of a given quantum program~\cite{Campbell17, Fowler12}. In contrast, IBM's quantum computers, which are examples of noisy-intermediate-scale-quantum (NISQ) systems, natively support the $U$ gate, $U(\theta, \psi, \lambda) = R_x(\psi)R_y(\theta)R_z(\lambda)$, which is parameterized over 3 continuous variables, and the $\mathrm{CNOT}$ gate. Usually, 2-qubit gates are more error-prone than single qubit gates. For this reason, a good measure for the cost of a quantum circuit synthesized for NISQ machines is the number of $\mathrm{CNOT}$ gates. The same is also valid for the Controlled-$Z$ gate that is used, e.g., in Rigetti's NISQ systems.

\subsection{Approximation Errors in Quantum Circuits}\label{sect:errorqc}

As mentioned in Section~\ref{sec:intro}, our framework addresses approximation errors, which may be reduced at an increased implementation cost. For the example of \emph{synthesis errors}, which occur due to the discrete nature of fault-tolerant instruction sets, the trade-off between the number of $T$ gates and the approximation error is logarithmic~\cite{KMM2013}. That is, to achieve an approximation error $\varepsilon$, $\mathcal O(\log(1/\varepsilon))$ $T$ gates are sufficient. In a typical quantum program, multiple different sources of such errors are present and thus the question arises: \textit{How do the approximation errors of individual operations compose?}

In this section, we prove that it is possible to derive an upper bound on the total approximation error by summing up all the individual approximation errors. Consequently, our methodology produces quantum circuits with accuracy guarantees for quantum programs without measurement feedback. Note that this excludes programs that rely on repeat-until-success statements, i.e., loops that iterate until a certain measurement outcome is observed (see, e.g., \citet{paetznick13}). Such cases can be handled separately using, e.g., an upper bound on the number of iterations.
In the last section, we propose a possible way of enabling accuracy management for such programs as future work. In all other cases, branching on measurement results may be addressed using an expression for the total error of the form $\varepsilon_{b}=\operatorname{max}(\varepsilon_1, \varepsilon_2)$, where $\varepsilon_1$ and $\varepsilon_2$ denote the errors of each branch. This follows from the \textit{Deferred Measurement Principle}~\cite[Section 4.4]{NC2000}, which says that measurements may be delayed until the end of the computation, transforming all quantum operations that are executed conditionally on the measurement result into quantum-controlled operations.

Quantum circuits corresponding to feedback-free programs consist of a sequence of gates $U_1, ..., U_m$, followed by a sequence of measurements $M_1, ..., M_k$ that produce a measurement outcome $m=x_i$ for a final state 
\[
\ket\psi = U_m\cdots U_1\ket{0}^{\otimes N}
\]
with probability
\[
	P(m=x_i) = |\braket{x_i | \psi}|^2,
\]
where $\braket{x_i|\psi}$ denotes the overlap of $\ket{\psi}$ with $\ket{x_i}$.

Therefore, our methodology must ensure that the actual final state $\ket{\tilde\psi}$ is close to the desired final state $\ket{\psi}$ after all decompositions have been applied to $U_1, ..., U_m$.

Let $V_1,...,V_n$ be an approximate decomposition of the quantum program in terms of the gates supported by the target hardware, i.e.,
\[
	\|\underbrace{U_m\cdots U_1}_U - \underbrace{V_n\cdots V_1}_V\| \leq \varepsilon,
\]
where $\|\cdot\|$ denotes the spectral norm as defined in~\citet[Section 2.1.4]{NC2000}.
Then, $\|\ket{\psi}-\ket{\tilde{\psi}}\| = \|U\ket{0}^{\otimes N}-V\ket{0}^{\otimes N}\|\leq\varepsilon$, which guarantees that, with $\ket\psi = \sum_i a_i\ket{x_i}$ and $\ket{\tilde\psi}=\sum_i\tilde a_i\ket{x_i}$,
\begin{align*}
	|a_i - \tilde a_i| &= \sqrt{|a_i - \tilde a_i|^2}\\
	&\leq \sqrt{\sum_i|a_i - \tilde a_i|^2}\\
	&\leq \varepsilon.
\end{align*}
Therefore, it is sufficient that our methodology guarantees
\[
\|U-V\|\leq\varepsilon.
\]
For more details on approximated unitary operators, we refer to Box 4.1 \emph{``Approximating quantum circuits"} of~\citet{NC2000}.

In the process of translating the quantum program to the native gate set, several decompositions are applied that introduce approximation errors. Let $U$ be a quantum operation being approximated by the decomposition into $W_1,...,W_t$. The decomposition introduces at most $\varepsilon_U$ if $\|U - (W_t\cdots W_1)\|\leq\varepsilon_U$, assuming that all $W_i$ are implemented exactly. Now, combining multiple such approximate implementations $\tilde U_i$ of $U_i$ such that $\|U_i-\tilde U_i\|\leq\varepsilon_i$ yields a total error of at most $\sum_i\varepsilon_i$~\cite{bernstein1997quantum}.

Therefore, it is possible to derive recursively-defined expressions for the error $E(U, \varepsilon_U)$ and the total gate count $T(U, \varepsilon_U)$~\cite{hrs2018}:
\begin{align*}
	E(U, \varepsilon_U) &= \varepsilon_U + \sum_{W\in D(U, \varepsilon_U)} E(W, \varepsilon_{W})f_W(\varepsilon_U),\\
	T(U, \varepsilon_U) &= \sum_{W\in D(U, \varepsilon_U)} T(W, \varepsilon_{W}) f_W(\varepsilon_U),
\end{align*}
where $D(U,\varepsilon_U)$ is the set of gates in the $\varepsilon_U$-approximate decomposition of $U$ and $f_W(\varepsilon_U)$ denotes the number of $W$ operations in the decomposition. Looking at these expressions, it is clear that an upper-bound on the total error can be computed very similarly to counting gates.

In conclusion, by applying this reasoning to the main entry point of a quantum program, we can choose all $\varepsilon_{(\cdot)}$ in the expression for
\[
	E(U_{Main},\varepsilon_{U_{Main}})
\]
such that $E(U_{Main}, \varepsilon_{U_{Main}}) \leq \varepsilon$. This ensures that the measurement probability amplitude for a given bit-string changes by at most $\varepsilon$.

\begin{figure*}[t]
\resizebox{.95\textwidth}{!}{%
\begin{tikzpicture}[scale=1.000000,x=1pt,y=1pt]
\filldraw[color=white] (0.000000, -7.500000) rectangle (405.000000, 67.500000);
\draw[color=black] (0.000000,60.000000) -- (405.000000,60.000000);
\draw[color=black] (0.000000,60.000000) node[left] {$|j_1\rangle$};
\draw[color=black] (0.000000,45.000000) -- (405.000000,45.000000);
\draw[color=black] (0.000000,45.000000) node[left] {$|j_2\rangle$};
\draw[color=black] (0.000000,30.000000) node[anchor=mid east] {$\vdots$};
\draw[color=black] (0.000000,15.000000) -- (405.000000,15.000000);
\draw[color=black] (0.000000,15.000000) node[left] {$|j_{n-1}\rangle$};
\draw[color=black] (0.000000,0.000000) -- (405.000000,0.000000);
\draw[color=black] (0.000000,0.000000) node[left] {$|j_n\rangle$};
\begin{scope}
\draw[fill=white] (12.000000, 60.000000) +(-45.000000:8.485281pt and 8.485281pt) -- +(45.000000:8.485281pt and 8.485281pt) -- +(135.000000:8.485281pt and 8.485281pt) -- +(225.000000:8.485281pt and 8.485281pt) -- cycle;
\clip (12.000000, 60.000000) +(-45.000000:8.485281pt and 8.485281pt) -- +(45.000000:8.485281pt and 8.485281pt) -- +(135.000000:8.485281pt and 8.485281pt) -- +(225.000000:8.485281pt and 8.485281pt) -- cycle;
\draw (12.000000, 60.000000) node {$H$};
\end{scope}
\draw (43.000000,60.000000) -- (43.000000,45.000000);
\begin{scope}
\draw[fill=white] (43.000000, 60.000000) +(-45.000000:18.384776pt and 8.485281pt) -- +(45.000000:18.384776pt and 8.485281pt) -- +(135.000000:18.384776pt and 8.485281pt) -- +(225.000000:18.384776pt and 8.485281pt) -- cycle;
\clip (43.000000, 60.000000) +(-45.000000:18.384776pt and 8.485281pt) -- +(45.000000:18.384776pt and 8.485281pt) -- +(135.000000:18.384776pt and 8.485281pt) -- +(225.000000:18.384776pt and 8.485281pt) -- cycle;
\draw (43.000000, 60.000000) node {$R_2$};
\end{scope}
\filldraw (43.000000, 45.000000) circle(1.500000pt);
\draw[color=black] (75.500000, 60.000000) node [fill=white] {$\cdots$};
\draw[color=black] (75.500000, 45.000000) node [fill=white] {$\cdots$};
\draw (108.000000,60.000000) -- (108.000000,15.000000);
\begin{scope}
\draw[fill=white] (108.000000, 60.000000) +(-45.000000:18.384776pt and 8.485281pt) -- +(45.000000:18.384776pt and 8.485281pt) -- +(135.000000:18.384776pt and 8.485281pt) -- +(225.000000:18.384776pt and 8.485281pt) -- cycle;
\clip (108.000000, 60.000000) +(-45.000000:18.384776pt and 8.485281pt) -- +(45.000000:18.384776pt and 8.485281pt) -- +(135.000000:18.384776pt and 8.485281pt) -- +(225.000000:18.384776pt and 8.485281pt) -- cycle;
\draw (108.000000, 60.000000) node {$R_{n-1}$};
\end{scope}
\filldraw (108.000000, 15.000000) circle(1.500000pt);
\draw (146.000000,60.000000) -- (146.000000,0.000000);
\begin{scope}
\draw[fill=white] (146.000000, 60.000000) +(-45.000000:18.384776pt and 8.485281pt) -- +(45.000000:18.384776pt and 8.485281pt) -- +(135.000000:18.384776pt and 8.485281pt) -- +(225.000000:18.384776pt and 8.485281pt) -- cycle;
\clip (146.000000, 60.000000) +(-45.000000:18.384776pt and 8.485281pt) -- +(45.000000:18.384776pt and 8.485281pt) -- +(135.000000:18.384776pt and 8.485281pt) -- +(225.000000:18.384776pt and 8.485281pt) -- cycle;
\draw (146.000000, 60.000000) node {$R_{n}$};
\end{scope}
\filldraw (146.000000, 0.000000) circle(1.500000pt);
\begin{scope}
\draw[fill=white] (177.000000, 45.000000) +(-45.000000:8.485281pt and 8.485281pt) -- +(45.000000:8.485281pt and 8.485281pt) -- +(135.000000:8.485281pt and 8.485281pt) -- +(225.000000:8.485281pt and 8.485281pt) -- cycle;
\clip (177.000000, 45.000000) +(-45.000000:8.485281pt and 8.485281pt) -- +(45.000000:8.485281pt and 8.485281pt) -- +(135.000000:8.485281pt and 8.485281pt) -- +(225.000000:8.485281pt and 8.485281pt) -- cycle;
\draw (177.000000, 45.000000) node {$H$};
\end{scope}
\draw[color=black] (202.500000, 45.000000) node [fill=white] {$\cdots$};
\draw (235.000000,45.000000) -- (235.000000,15.000000);
\begin{scope}
\draw[fill=white] (235.000000, 45.000000) +(-45.000000:18.384776pt and 8.485281pt) -- +(45.000000:18.384776pt and 8.485281pt) -- +(135.000000:18.384776pt and 8.485281pt) -- +(225.000000:18.384776pt and 8.485281pt) -- cycle;
\clip (235.000000, 45.000000) +(-45.000000:18.384776pt and 8.485281pt) -- +(45.000000:18.384776pt and 8.485281pt) -- +(135.000000:18.384776pt and 8.485281pt) -- +(225.000000:18.384776pt and 8.485281pt) -- cycle;
\draw (235.000000, 45.000000) node {$R_{n-2}$};
\end{scope}
\filldraw (235.000000, 15.000000) circle(1.500000pt);
\draw (273.000000,45.000000) -- (273.000000,0.000000);
\begin{scope}
\draw[fill=white] (273.000000, 45.000000) +(-45.000000:18.384776pt and 8.485281pt) -- +(45.000000:18.384776pt and 8.485281pt) -- +(135.000000:18.384776pt and 8.485281pt) -- +(225.000000:18.384776pt and 8.485281pt) -- cycle;
\clip (273.000000, 45.000000) +(-45.000000:18.384776pt and 8.485281pt) -- +(45.000000:18.384776pt and 8.485281pt) -- +(135.000000:18.384776pt and 8.485281pt) -- +(225.000000:18.384776pt and 8.485281pt) -- cycle;
\draw (273.000000, 45.000000) node {$R_{n-1}$};
\end{scope}
\filldraw (273.000000, 0.000000) circle(1.500000pt);
\draw[color=black] (305.500000, 45.000000) node [fill=white] {$\cdots$};
\draw[color=black] (305.500000, 15.000000) node [fill=white] {$\cdots$};
\draw[color=black] (305.500000, 0.000000) node [fill=white] {$\cdots$};
\begin{scope}
\draw[fill=white] (331.000000, 15.000000) +(-45.000000:8.485281pt and 8.485281pt) -- +(45.000000:8.485281pt and 8.485281pt) -- +(135.000000:8.485281pt and 8.485281pt) -- +(225.000000:8.485281pt and 8.485281pt) -- cycle;
\clip (331.000000, 15.000000) +(-45.000000:8.485281pt and 8.485281pt) -- +(45.000000:8.485281pt and 8.485281pt) -- +(135.000000:8.485281pt and 8.485281pt) -- +(225.000000:8.485281pt and 8.485281pt) -- cycle;
\draw (331.000000, 15.000000) node {$H$};
\end{scope}
\draw (362.000000,15.000000) -- (362.000000,0.000000);
\begin{scope}
\draw[fill=white] (362.000000, 15.000000) +(-45.000000:18.384776pt and 8.485281pt) -- +(45.000000:18.384776pt and 8.485281pt) -- +(135.000000:18.384776pt and 8.485281pt) -- +(225.000000:18.384776pt and 8.485281pt) -- cycle;
\clip (362.000000, 15.000000) +(-45.000000:18.384776pt and 8.485281pt) -- +(45.000000:18.384776pt and 8.485281pt) -- +(135.000000:18.384776pt and 8.485281pt) -- +(225.000000:18.384776pt and 8.485281pt) -- cycle;
\draw (362.000000, 15.000000) node {$R_2$};
\end{scope}
\filldraw (362.000000, 0.000000) circle(1.500000pt);
\begin{scope}
\draw[fill=white] (393.000000, -0.000000) +(-45.000000:8.485281pt and 8.485281pt) -- +(45.000000:8.485281pt and 8.485281pt) -- +(135.000000:8.485281pt and 8.485281pt) -- +(225.000000:8.485281pt and 8.485281pt) -- cycle;
\clip (393.000000, -0.000000) +(-45.000000:8.485281pt and 8.485281pt) -- +(45.000000:8.485281pt and 8.485281pt) -- +(135.000000:8.485281pt and 8.485281pt) -- +(225.000000:8.485281pt and 8.485281pt) -- cycle;
\draw (393.000000, -0.000000) node {$H$};
\end{scope}
\draw[color=black] (405.000000,60.000000) node[right] {$|0\rangle+e^{2 \pi i 0.j_1\dots j_n}|1\rangle$};
\draw[color=black] (405.000000,45.000000) node[right] {$|0\rangle+e^{2 \pi i 0.j_2\dots j_n}|1\rangle$};
\draw[color=black] (405.000000,30.000000) node[anchor=mid west] {$\vdots$};
\draw[color=black] (405.000000,15.000000) node[right] {$|0\rangle+e^{2 \pi i 0.j_{n-1} j_n}|1\rangle$};
\draw[color=black] (405.000000,0.000000) node[right] {$|0\rangle+e^{2 \pi i 0.j_n}|1\rangle$};
\end{tikzpicture}
}
\caption{Efficient circuit computing the quantum Fourier transform.}
\label{figure:qft}
\end{figure*}
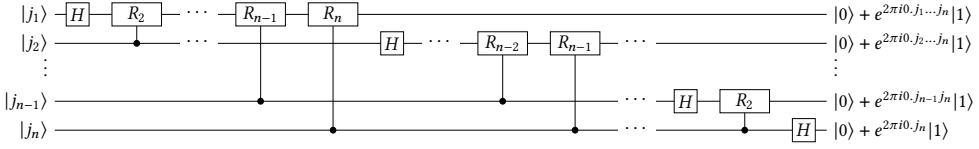

\section{Sample Quantum Programs}\label{sec:sample}

\subsection{Quantum Fourier Transform}\label{sec:qft}
The quantum Fourier transform (QFT) is an algorithm that performs the Fourier transform of quantum mechanical amplitudes. QFT is implemented as a linear operator that applies the following unitary transformation on a basis state $|j\rangle$~\cite{NC2000}:
\[
|j\rangle \mapsto \frac{1}{\sqrt{N}}\sum_{k=0}^{N-1}e^{2\pi i j k /N}|k\rangle
\]
The effect of the transform on an arbitrary state can be described using a product representation that maps $|j_1, \dots, j_n\rangle$ to
\begin{small}
\[
\frac{
(|0\rangle+e^{2\pi i 0.j_n}|1\rangle) (|0\rangle+e^{2\pi i 0.j_{n-1}j_n}|1\rangle) \cdots 
(|0\rangle+e^{2\pi i 0.j_1j_2\dots j_n}|1\rangle) 
}{2^{n/2}}
\]
\end{small}%
where $0.j_lj_{l+1}\dots j_n $ is the binary expansion $j_l/2+j_{l+1}/4+ \dots + j_n/2^{n-l+1}$.
This representation has a direct correspondence to the circuit implementation of the quantum Fourier transform shown in Fig.~\ref{figure:qft}. 
The circuit is composed of $n$ steps, one for each qubit. In each step, for each qubit, a Hadamard gate is applied, followed by a series of rotation gates controlled by all remaining qubits. 

In this work, we will largely refer to an approximate version of QFT (AQFT) in which the number of rotations is reduced according to the desired approximation error $\varepsilon_{QFT}$. This is done by pruning the rotations with small angles. In particular, for each qubit $j_i$ with $1\leq i < n$ a maximum of
\[
l = \lceil \log_2(n / \varepsilon_{QFT})\rceil+3
\]
controlled-rotations is applied~\cite{coppersmith02}. 

The quantum Fourier transform enables the \emph{quantum phase estimation} algorithm and has a key role in the solution of many relevant problems, e.g., the \emph{integer factorization} problem.

\subsection{Simulating Time-Evolution of Operators}

Being a quantum system, a quantum computer can be programmed to simulate other quantum systems. This, for example, can be used to elucidate chemical reaction mechanisms~\cite{reiher17}.

Given the quantum-mechanical Hamiltonian $\mathcal H$ which describes the system being studied, the time evolution of the system can be simulated by implementing the \textit{time-evolution operator} $U=e^{-i\mathcal Ht}$. The time-evolved quantum state can then be obtained by applying $U$ to the initial state $\ket{\psi(0)}$:
\[
|\psi(t)\rangle = e^{-i\mathcal H t}|\psi(0)\rangle
\]
In order to implement the time-evolution operator on a quantum computer, it needs to be decomposed into the native gate set, e.g., Clifford+$T$. Different decomposition methods are available, each one with a different scaling with respect to the targeted precision $\varepsilon_{TE}$: polynomial for the Trotter decomposition method, logarithmic for the linear combination of unitaries (LCU)~\cite{babbush17, poulin14}.

\begin{example}
Consider the Hamiltonian of a 1D transverse-field Ising model (TFIM)
\[
 \mathcal H = \underbrace{-J\sum_{\langle i,j \rangle}Z^{i}Z^{j}}_{\mathcal H_1} - \underbrace{h\sum_i X^{i}}_{\mathcal H_2},
\]
where $\mathcal H_1$ defines the interaction of adjacent spins, denoted by $\langle i,j\rangle$, with periodic boundary conditions, $\mathcal H_2$ defines the interaction of the system with the external transverse field, and $X^i, Z^i$ denote the application of $X=\begin{psmallmatrix}0 & 1\\1 & 0\end{psmallmatrix}$ and $Z=\begin{psmallmatrix}1 & 0\\0 & -1\end{psmallmatrix}$, respectively, to spin $i$.

Using a second-order Trotter-Suzuki decomposition, the time-evolution operator under this Hamiltonian can be written as
\[
	e^{-i\mathcal Ht} \approx (e^{-i\mathcal H_1\frac{t}{2M}} e^{i\mathcal H_2\frac{t}{M}} e^{-i\mathcal H_1\frac{t}{2M}})^M.
\]
The number of Trotter steps $M$ will be chosen according to the desired accuracy $\varepsilon_{TE}$.
In particular, for this second-order Trotter decomposition we have that $M$ is proportional to $1/\sqrt{\varepsilon_{TE}}$~\cite{reiher17}. 
Each Trotter step can be implemented using $\mathrm{CNOT}$ and $R_Z(\theta)$ gates. The latter being a gate that applies a rotation equal to the angle $\theta$ around the z-axis. 
Considering Clifford+$T$ as the native gate set, each rotation has to be \textit{synthesized} or decomposed in terms of these gates. As not every rotation can be realized exactly using this gate library, rotation synthesis also introduces an error.
Given a target approximation error $\varepsilon_{R}$, the number of $T$ gates per rotation will be proportional to $\log_2{\left(\frac1{\varepsilon_R}\right)}$.

Thus, to express the time-evolution operator we need to take into account two inter-dependent approximation errors, namely $\varepsilon_{TE}$ and $\varepsilon_{R}$.
\end{example}

\subsection{Quantum Phase Estimation}
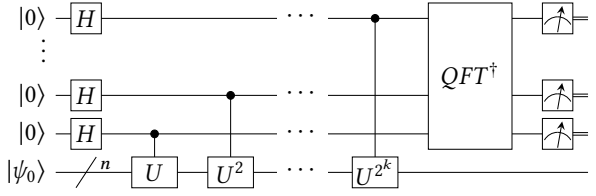
\begin{figure}[t]
\resizebox{.6\textwidth}{!}{%
\begin{tikzpicture}[scale=1.000000,x=1pt,y=1pt]
\filldraw[color=white] (0.000000, -7.500000) rectangle (210.000000, 67.500000);
\draw[color=black] (0.000000,60.000000) -- (198.000000,60.000000);
\draw[color=black] (198.000000,59.500000) -- (210.000000,59.500000);
\draw[color=black] (198.000000,60.500000) -- (210.000000,60.500000);
\draw[color=black] (0.000000,60.000000) node[left] {$|0\rangle$};
\draw[color=black] (0.000000,45.000000) node[anchor=mid east] {$\vdots$};
\draw[color=black] (0.000000,30.000000) -- (198.000000,30.000000);
\draw[color=black] (198.000000,29.500000) -- (210.000000,29.500000);
\draw[color=black] (198.000000,30.500000) -- (210.000000,30.500000);
\draw[color=black] (0.000000,30.000000) node[left] {$|0\rangle$};
\draw[color=black] (0.000000,15.000000) -- (198.000000,15.000000);
\draw[color=black] (198.000000,14.500000) -- (210.000000,14.500000);
\draw[color=black] (198.000000,15.500000) -- (210.000000,15.500000);
\draw[color=black] (0.000000,15.000000) node[left] {$|0\rangle$};
\draw[color=black] (0.000000,0.000000) -- (210.000000,0.000000);
\draw[color=black] (0.000000,0.000000) node[left] {$|\psi_0\rangle$};
\draw (8.000000, -6.000000) -- (16.000000, 6.000000);
\draw (14.000000, 3.000000) node[right] {$\scriptstyle{n}$};
\begin{scope}
\draw[fill=white] (12.000000, 60.000000) +(-45.000000:8.485281pt and 8.485281pt) -- +(45.000000:8.485281pt and 8.485281pt) -- +(135.000000:8.485281pt and 8.485281pt) -- +(225.000000:8.485281pt and 8.485281pt) -- cycle;
\clip (12.000000, 60.000000) +(-45.000000:8.485281pt and 8.485281pt) -- +(45.000000:8.485281pt and 8.485281pt) -- +(135.000000:8.485281pt and 8.485281pt) -- +(225.000000:8.485281pt and 8.485281pt) -- cycle;
\draw (12.000000, 60.000000) node {$H$};
\end{scope}
\begin{scope}
\draw[fill=white] (12.000000, 30.000000) +(-45.000000:8.485281pt and 8.485281pt) -- +(45.000000:8.485281pt and 8.485281pt) -- +(135.000000:8.485281pt and 8.485281pt) -- +(225.000000:8.485281pt and 8.485281pt) -- cycle;
\clip (12.000000, 30.000000) +(-45.000000:8.485281pt and 8.485281pt) -- +(45.000000:8.485281pt and 8.485281pt) -- +(135.000000:8.485281pt and 8.485281pt) -- +(225.000000:8.485281pt and 8.485281pt) -- cycle;
\draw (12.000000, 30.000000) node {$H$};
\end{scope}
\begin{scope}
\draw[fill=white] (12.000000, 15.000000) +(-45.000000:8.485281pt and 8.485281pt) -- +(45.000000:8.485281pt and 8.485281pt) -- +(135.000000:8.485281pt and 8.485281pt) -- +(225.000000:8.485281pt and 8.485281pt) -- cycle;
\clip (12.000000, 15.000000) +(-45.000000:8.485281pt and 8.485281pt) -- +(45.000000:8.485281pt and 8.485281pt) -- +(135.000000:8.485281pt and 8.485281pt) -- +(225.000000:8.485281pt and 8.485281pt) -- cycle;
\draw (12.000000, 15.000000) node {$H$};
\end{scope}
\draw (39.000000,15.000000) -- (39.000000,0.000000);
\begin{scope}
\draw[fill=white] (39.000000, -0.000000) +(-45.000000:12.727922pt and 8.485281pt) -- +(45.000000:12.727922pt and 8.485281pt) -- +(135.000000:12.727922pt and 8.485281pt) -- +(225.000000:12.727922pt and 8.485281pt) -- cycle;
\clip (39.000000, -0.000000) +(-45.000000:12.727922pt and 8.485281pt) -- +(45.000000:12.727922pt and 8.485281pt) -- +(135.000000:12.727922pt and 8.485281pt) -- +(225.000000:12.727922pt and 8.485281pt) -- cycle;
\draw (39.000000, -0.000000) node {$U$};
\end{scope}
\filldraw (39.000000, 15.000000) circle(1.500000pt);
\draw (69.000000,30.000000) -- (69.000000,0.000000);
\begin{scope}
\draw[fill=white] (69.000000, -0.000000) +(-45.000000:12.727922pt and 8.485281pt) -- +(45.000000:12.727922pt and 8.485281pt) -- +(135.000000:12.727922pt and 8.485281pt) -- +(225.000000:12.727922pt and 8.485281pt) -- cycle;
\clip (69.000000, -0.000000) +(-45.000000:12.727922pt and 8.485281pt) -- +(45.000000:12.727922pt and 8.485281pt) -- +(135.000000:12.727922pt and 8.485281pt) -- +(225.000000:12.727922pt and 8.485281pt) -- cycle;
\draw (69.000000, -0.000000) node {$U^2$};
\end{scope}
\filldraw (69.000000, 30.000000) circle(1.500000pt);
\draw[color=black] (97.500000, 60.000000) node [fill=white] {$\cdots$};
\draw[color=black] (97.500000, 30.000000) node [fill=white] {$\cdots$};
\draw[color=black] (97.500000, 15.000000) node [fill=white] {$\cdots$};
\draw[color=black] (97.500000, 0.000000) node [fill=white] {$\cdots$};
\draw (126.000000,60.000000) -- (126.000000,0.000000);
\begin{scope}
\draw[fill=white] (126.000000, -0.000000) +(-45.000000:12.727922pt and 8.485281pt) -- +(45.000000:12.727922pt and 8.485281pt) -- +(135.000000:12.727922pt and 8.485281pt) -- +(225.000000:12.727922pt and 8.485281pt) -- cycle;
\clip (126.000000, -0.000000) +(-45.000000:12.727922pt and 8.485281pt) -- +(45.000000:12.727922pt and 8.485281pt) -- +(135.000000:12.727922pt and 8.485281pt) -- +(225.000000:12.727922pt and 8.485281pt) -- cycle;
\draw (126.000000, -0.000000) node {$U^{2^k}$};
\end{scope}
\filldraw (126.000000, 60.000000) circle(1.500000pt);
\draw (163.500000,60.000000) -- (163.500000,15.000000);
\begin{scope}
\draw[fill=white] (163.500000, 37.500000) +(-45.000000:23.334524pt and 40.305087pt) -- +(45.000000:23.334524pt and 40.305087pt) -- +(135.000000:23.334524pt and 40.305087pt) -- +(225.000000:23.334524pt and 40.305087pt) -- cycle;
\clip (163.500000, 37.500000) +(-45.000000:23.334524pt and 40.305087pt) -- +(45.000000:23.334524pt and 40.305087pt) -- +(135.000000:23.334524pt and 40.305087pt) -- +(225.000000:23.334524pt and 40.305087pt) -- cycle;
\draw (163.500000, 37.500000) node {$QFT^{\dagger}$};
\end{scope}
\draw[fill=white] (192.000000, 54.000000) rectangle (204.000000, 66.000000);
\draw[very thin] (198.000000, 60.600000) arc (90:150:6.000000pt);
\draw[very thin] (198.000000, 60.600000) arc (90:30:6.000000pt);
\draw[->,>=stealth] (198.000000, 54.600000) -- +(80:10.392305pt);
\draw[fill=white] (192.000000, 24.000000) rectangle (204.000000, 36.000000);
\draw[very thin] (198.000000, 30.600000) arc (90:150:6.000000pt);
\draw[very thin] (198.000000, 30.600000) arc (90:30:6.000000pt);
\draw[->,>=stealth] (198.000000, 24.600000) -- +(80:10.392305pt);
\draw[fill=white] (192.000000, 9.000000) rectangle (204.000000, 21.000000);
\draw[very thin] (198.000000, 15.600000) arc (90:150:6.000000pt);
\draw[very thin] (198.000000, 15.600000) arc (90:30:6.000000pt);
\draw[->,>=stealth] (198.000000, 9.600000) -- +(80:10.392305pt);
\draw[color=black] (210.000000,45.000000) node[anchor=mid west] {$\vdots$};
\end{tikzpicture}
}
\caption{Quantum circuit performing quantum-phase estimation on an $n$-qubit system with an accuracy of $k+1$ bits.}
\label{qca}
\end{figure}
Once time-evolution under the Hamiltonian $\mathcal H$ is implemented, one may perform measurements similar to experiments with the actual system. Quantum computing, however, allows us to achieve a quadratic advantage over repeated measurement and sampling via quantum phase estimation (QPE). 
Given a state with large overlap with the ground state $\ket{\psi_0}$ of the Hamiltonian, this algorithm allows us to determine the ground state energy $E_0$
\[
\mathcal H|\psi_0\rangle = E_0|\psi_0\rangle.
\]
Fig.~\ref{qca} shows one of several possible implementations of QPE~\cite{NC2000}. The measurement outcomes of the top $k+1$ qubits yield a $k+1$-bit approximation to the phase due to time-evolution. More precisely, the number of qubits to choose $n_{QPE}$ depends on the desired accuracy \textit{and} the probability $p$ of a successful measurement as
\[
n_{QPE} = n + \left\lceil \log\left(2 + \frac{1}{2(1-p)}\right)\right\rceil,
\]
where $n$ is the desired accuracy in number of bits.

The number of controlled time-evolution unitaries required for QPE to succeed with $p=0.5$ and accuracy $\varepsilon_{QPE}$ may thus be bounded by $2^{n_{QPE}}-1 \leq 16\pi /\varepsilon_{QPE}$.

Not only does QPE allow one to infer the ground state energy if the ground state is known, but it also collapses a non-eigenstate input $\ket{\phi}$ to the $i$-th eigenstate $\ket{\psi_i}$ of the Hamiltonian $H$ with probability $|\braket{\psi_i|\phi}|^2$.

In terms of accuracy, it is important to distinguish between the different applications of QPE. If QPE is used to determine only the energy, i.e., $|E_0-\tilde{E_0}|\leq \varepsilon$ is required, then it is suf\-fi\-cient to implement time-evolution such that $\|U-\tilde U\| \leq \varepsilon-\varepsilon_{QPE}$. However, if the goal is to prepare the ground state, i.e., $\| \ket{\psi_0} - \ket{\tilde{\psi_0}} \| \leq \varepsilon$, then $\|U-\tilde U\| \leq \frac{\varepsilon-\varepsilon_{QPE}}{2^{n_{QPE}}-1}$ is sufficient (both via triangle inequality). Distinguishing these cases clearly has a great impact on the resulting resource requirements.

\section{Adding Language Support for Accuracy Management}\label{sect:support}

Since large-scale quantum computers are not yet available, resource estimation is a crucial feature of any software framework for quantum computing.
Typically,  resource estimation is performed by compiling the quantum program into the chosen target gate set and then executing the resulting circuit on a classical simulator that counts native operations (instead of executing them).
For this to be possible, however, all the parameters of the program, including accuracy parameters for each subroutine, must be determined.

Existing quantum programming languages do not offer built-in support for accuracy management. Consequently, it is very cumbersome to implement large-scale quantum algorithms in an accuracy-aware fashion.
Thus, despite the availability of a wide range of quantum programming languages, resource estimates are still computed (semi-)manually, taking care of accuracy parameters using pen and paper~\cite{Scherer17, reiher17}.

The main difficulty when selecting appropriate accuracy parameters is that parameters at a higher level of abstraction have an effect on the ones at lower levels, as illustrated by the following example:
\begin{example}
	Consider QPE on $U=R_Z(\alpha):=e^{-i\frac\alpha2 Z}$ and the target gate set Clifford+$T$. The number of phase-estimation qubits $n_{QPE}$ depends on the desired precision of the phase (and the probability of success). At the lowest level, the various $U^{2^i}=R_Z(\alpha_i)$ are decomposed into a sequence of Clifford+$T$ gates featuring $O(\log \frac1{\varepsilon_r})$ $T$ gates, where $\varepsilon_r$ is the error of a single rotation. 
	To achieve an overall target accuracy $\varepsilon$, $\varepsilon_r$ must be chosen such that $\varepsilon_{QPE}+\varepsilon_R \leq \varepsilon$, where $\varepsilon_R$ denotes the error introduced by all rotations in the quantum circuit.
	 If the same accuracy is chosen for all the rotations, then $\epsilon_R = N_{rot} \epsilon_r$, where $N_{rot}$ is the number of rotations. 
	 Since $\varepsilon_{QPE}$ affects the number of rotations in the circuit, $\varepsilon_r$ must be chosen as a function of $\varepsilon_{QPE}$.
\end{example}
In general, it would be possible to adapt all values in the code manually on a case-by-case basis---however, this defeats the purpose of having a high-level programming language.
The lack of language support forces programmers to manually handle accuracy parameters by passing all such parameters to the main routine, which forwards them to each subroutine. In case of the QPE algorithm, this might result in code as follows: 

\begin{small}
\begin{table}[h!]
\begin{minipage}{.5\textwidth}
\begin{algorithmic}
\Function{QPE}{$\varepsilon_{QPE}$,  $\varepsilon_{TE}$,$\varepsilon_{QFT}$, $\varepsilon_{R_{TE}}$, $\varepsilon_{R_{QFT}}$, U}
  		\State $reg\_size \gets f(\varepsilon_{QPE})$
        \For{$i \gets 0$ to $reg\_size$ }
        		\For{$j \gets0$ to $n\_iter(i)$}
        		\State $^c$U($\varepsilon_{TE}$, $\varepsilon_{R_{TE}}$)
       		 \EndFor
      	\EndFor
        \State AQFT$^\dagger$($\varepsilon_{QFT}$, $\varepsilon_{R_{QFT}}$)
\EndFunction
\end{algorithmic}
\end{minipage}
\end{table}

\end{small}
\noindent
Here, $^cU$ is the controlled version of $U$; we also omitted all qubit variables for better readability.

This programmer-unfriendly approach does not allow code reuse for resource estimation, as implementations of subroutines need to be adapted to the context in which they are used and, in particular, to the choice of accuracy parameters. For example, $^c$U($\varepsilon_{TE}$, $\varepsilon_{R_{TE}}$) is implemented using a number of Clifford+$T$ gates that depends on the chosen accuracy parameters.

When using our methodology, programmers need to worry about accuracy parameters only in subroutines where the corresponding errors are introduced. The compiler will take care of extracting all dependencies. Specifically, this allows us to express the pseudo-code for phase estimation as follows: 
\begin{table}[H]
\centering
\begin{tabular}{c|c}
\begin{minipage}{.3\textwidth}
\small
\begin{algorithmic}
\Function{$^c$R}{}
	\State declare $\varepsilon_{R}$
	\State $\dots$
\EndFunction
\Function{$^c$U}{}
	\State declare $\varepsilon_{TE}$
	\State $\dots$
\EndFunction
\Function{AQFT$^\dagger$}{}
	\State declare $\varepsilon_{QFT}$
	\State $l \gets g(\varepsilon_{QFT})$
	\For {$i \gets 0$ to $g'(l)$}
		\State $\dots$
		\State $^c$R()
		\State $\dots$
	\EndFor
\EndFunction
\end{algorithmic}
\end{minipage} & 
\begin{minipage}{.35\textwidth}
\small
\begin{algorithmic}
\Function{QPE}{U}
\State declare $\varepsilon_{QPE}$
  		\State $reg\_size \gets f(\varepsilon_{QPE})$
        \For{$i \gets 0$ to $reg\_size$ }
        		\For{$j \gets0$ to $n\_iter(i)$}
        		\State $^c$U()
       		 \EndFor
      	\EndFor
        \State AQFT$^\dagger$()
        \State
        \State
        \State
        \State
        \State
        \State
\EndFunction
\end{algorithmic}
\end{minipage}\\
\end{tabular}
\end{table}
Using abstract syntax tree (AST) transformations, our methodology is able to handle various levels of granularity: from using the same value for all accuracy parameters to using a different value for every instance that is created during runtime (via an accuracy parameter data structure that mirrors the call graph). This is crucial as there is a substantial trade-off between the number of accuracy parameters being considered and the resulting gate count~\cite{hrs2018}. This can be illustrated with the following example:

\begin{example}
Consider Beauregard's implementation of Shor's algorithm~\cite{beauregard02}. In addition to the (semi-classical) inverse Fourier transform of phase estimation, every addition circuit requires two (approximate) QFTs~\cite{coppersmith02} (one inverted, one regular)~\cite{draper00}. While the number of additions (and thus the number of QFTs) varies with the bit-size $n$ of the number to factor, phase estimation always requires a single QFT. Therefore, it is natural to choose a different accuracy parameter for the (approximate) QFT of the phase estimation than for the (approximate) QFTs of the $\mathcal O(n^2)$-many $n$-bit additions.
\end{example}

Besides facilitating accuracy-aware implementations of quantum programs and providing various levels of granularity for assigning accuracy parameters, our methodology allows us to automatically deduce the number of contexts in which a given (approximate) decomposition is applied. This enables automatic selection of the number of accuracy parameters and thus removes the need to perform this task manually.

\section{Automating Accuracy Management}\label{sec:automating}
\begin{figure}[t]
\definecolor{myblue}{RGB}{0, 102, 153}
\begin{tikzpicture}
\begin{scope}[every node/.style={rectangle, draw=black, minimum size=.55cm}]
  \node (qpe) [] {QPE};

  \node (cu2) [below =2cm of qpe] {$^cU$};
  \node (rcu21) [below right=1.5cm and .3cm of cu2.center] {R};
  \node (rcu22) [below left =1.5cm and .3cm of cu2.center] {R};

  \node (cu1) [left=1.5cm of cu2] {$^cU$};
  \node (rcu11) [below right=1.5cm and .3cm of cu1.center] {R};
  \node (rcu12) [below left =1.5cm and .3cm of cu1.center] {R};

  \node (qft) [right=1.5cm of cu2] {AQFT${}^\dagger$};
  \node (rqft1) [below right=1.5cm and .3cm of qft.center] {R};
  \node (rqft2) [below left =1.5cm and .3cm of qft.center] {R};

\end{scope}

\begin{scope}[every node/.style={minimum size=.15cm, inner sep=0pt}]
  \node (dot1) [below=1.5cm of cu1] {$\dots$};
  \node (dot2) [below=1.5cm of cu2] {$\dots$};  
  \node (dot3) [below=1.5cm of qft] {$\dots$};
  \node (dot4) [left=.75cm of cu2.center] {$\dots$};
\end{scope}

\begin{scope}[every node/.style={circle, draw=myblue, inner sep=0pt, outer sep=0pt, align=center}]
  \node (s1) [below right=.5cm and .75cm of cu1.center] {\textcolor{myblue}{$+$}};
  \node (s2) [below right=.5cm and .75cm of cu2.center] {\textcolor{myblue}{$+$}};
  \node (s3) [below right=.5cm and .75cm of qft.center] {\textcolor{myblue}{$+$}};

  \node (s5) [above right=1cm and 1.25cm of cu1.center] {\textcolor{myblue}{$+$}};
  \node (s6) [above right=1cm and 1.25cm of cu2.center] {\textcolor{myblue}{$+$}};
  \node (s7) [above right=1cm and 1.25cm of qft.center] {\textcolor{myblue}{$+$}};

  \node (s4) [right=1.5cm of qpe] {\textcolor{myblue}{$+$}};

\end{scope}
  \node (out) [above=.2cm of s4] {\textcolor{myblue}{$\varepsilon$}};

\path[->, color=myblue] (qpe) edge[line width=0.2mm] node[ fill=white, anchor=center, pos=0.5,font=\bfseries] {$\varepsilon_{QPE}$}(s4);
\path[->, color=myblue] (rcu11) edge[line width=0.2mm] node[ fill=white, anchor=center, pos=0.5,font=\bfseries] {$\varepsilon_{R}$}(s1);
\path[->, color=myblue] (cu2) edge[line width=0.2mm] node[ fill=white, anchor=center, pos=0.5,font=\bfseries] {$\varepsilon_{^cU}$}(s6);
\path[->, color=myblue] (qft)edge[line width=0.2mm] node[ fill=white, anchor=center, pos=0.5,font=\bfseries] {$\varepsilon_{QFT}$}(s7);

\draw[->, color=myblue] (rcu12)--(s1);
\draw[->, color=myblue] (rcu21)--(s2);
\draw[->, color=myblue] (rcu22)--(s2);
\draw[->, color=myblue] (rqft1)--(s3);
\draw[->, color=myblue] (rqft2)--(s3);
\draw[->, color=myblue] (cu1)--(s5);
\draw[->, color=myblue] (s3)--(s7);
\draw[->, color=myblue] (s2)--(s6);
\draw[->, color=myblue] (s1)--(s5);
\draw[->, color=myblue] (s4)--(out);
\draw[->, color=myblue] (s5)--(s4);
\draw[->, color=myblue] (s6)--(s4);
\draw[->, color=myblue] (s7)--(s4);

\draw[->] (qpe)--(cu2);
\draw[->] (qpe)--(cu1);
\draw[->] (qpe)--(qft);
\draw[->] (cu1)--(rcu11);
\draw[->] (cu1)--(rcu12);
\draw[->] (cu2)--(rcu21);
\draw[->] (cu2)--(rcu22);
\draw[->] (qft)--(rqft1);
\draw[->] (qft)--(rqft2);

\end{tikzpicture}
\caption{Flow diagram explaining how the code of the QPE algorithm is transformed into a code evaluating the overall approximation error $\varepsilon$.}
\label{sch}
\end{figure}
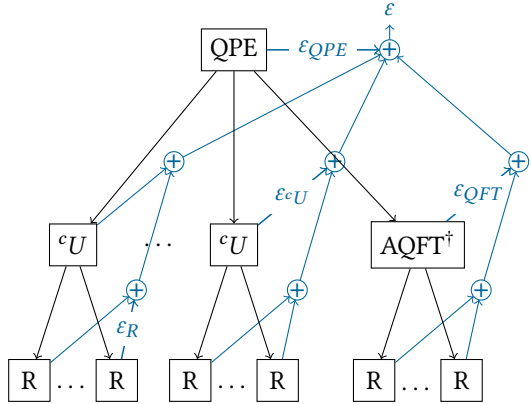

In this section, we describe the proposed procedure to automatically determine accuracy parameters. The optimization problem is solved using a simulated annealing procedure that iteratively changes the parameters and evaluates the corresponding total approximation error. The procedure terminates as soon as accuracy parameters have been found that guarantee a user-specified overall accuracy.
To further improve the parameter selection, we also evaluate the circuit cost in terms of number of expensive gates and pass the information to the optimization procedure. The result is a valid distribution of the available approximation error that in addition aims to minimize the circuit cost.

Our approach is to extract a (near-)symbolic expression for the total error and gate count from the algorithm and to use the obtained expressions in the annealing procedure. In general, it is possible to use the annealing procedure with any available resource estimation method, the difference being that the runtime of each evaluation depends on the problem size if non-symbolic methods are used.

In the following sections, we use the $T$-count as a measure of the implementation cost of a quantum circuit, as this closely captures the cost in a fault-tolerant setting. We also envision that our framework is most beneficial in this setting, since larger programs lead to larger improvements over non-symbolic approaches. Nevertheless, we note that our approach may be employed with any cost function. For example, one may want to select accuracy parameters while aiming to minimize the number of $\mathrm{CNOT}$ or $\mathrm{CZ}$ gates when targeting NISQ devices.

\subsection{Cost/Constraint Functions: Extraction}
Our methodology proceeds by automatically generating two pieces of code that compute (an upper bound on) (1) the number of costly quantum gates (T) and (2) the overall approximation error as a function of the different approximation errors (E). We denote the two functions by
\[
T(\varepsilon_1, \dots, \varepsilon_n) \qquad \text{and} \qquad E(\varepsilon_1, \dots, \varepsilon_n).
\]
The automatic generation of these two functions can be achieved via a few simple transformations of the program's AST. Specifically, to generate $T$, all calls to native operations are removed from the AST, except those corresponding to costly gates that are replaced by counter-increments. The program computing a bound on the overall approximation error ($E$) can be generated in a similar fashion, where increments are added for every epsilon declaration (see Sec.~\ref{sect:errorqc}).

\begin{example}
In Fig.~\ref{sch} we show the different decomposition levels of the QPE algorithm. The standard coherent QPE requires $2^{n_{QPE}}-1$ controlled time-evolution unitaries $^cU$, followed by an inverse QFT. At the next decomposition level, each unitary (including the QFT) is decomposed into rotation gates. In turn, those rotations will be fed into rotation synthesis, which out\-puts a sequence of $\mathcal O(\log\frac1{\varepsilon_R})$ Clifford+$T$ gates for each rotation, where $\varepsilon_R$ denotes the target accuracy of rotation synthesis (per rotation). Sin\-ce errors accumulate at most linearly due to being unitary (see Sec.~\ref{sect:errorqc}), an upper bound on the overall approximation error can be computed by adding all the $\varepsilon_i$ introduced by the various decomposition steps.
\end{example}

\subsection{Cost/Constraint Functions: Optimization}

Once the two pieces of code evaluating the total approximation error $E(\varepsilon_1,...,\varepsilon_n)$ and the cost $T(\varepsilon_1,...,\varepsilon_n)$  have been generated, they could be fed into the simulated annealing procedure.
While this would allow us to perform accuracy management \textit{automatically}, the resulting code will take substantial time to execute: typical quantum applications require on the order of $10^{15}$ or more operations~\cite{reiher17} and the optimization loop is executed hundreds of times until suitable accuracy parameters are found.

As a remedy, we employ custom compiler optimization passes that significantly reduce the time required to evaluate gate counts and error bounds.
Specifically, our methodology aims to infer symbolic and loop-free expressions for (upper bounds on) gate count and overall approximation error. 
The following example demonstrates how beneficial the use of our symbolic approach is:

\begin{example}
Consider the example of the approximate quantum phase estimation algorithm and a two-mode simulated annealing procedure. Even with an optimized annealing schedule, it will require a minimum of about 200 evaluations to guarantee an overall approximation error of at most $10^{-2}$. As accuracy parameters $\varepsilon_1 \dots \varepsilon_n$ approach the optimal values (minimizing the $T$-count), one evaluation of the non-optimized $T(\varepsilon_1,...,\varepsilon_n)$ function on 8 qubits takes $9m\, 10s$, while evaluating the inferred symbolic expression takes $0.1 \mu s$. If the number of qubits grows to 16, then we have $ 34m\, 14s $ for the non-optimized case, while evaluating the expression still takes  $0.1 \mu s$.
\end{example}

The transformations that we propose to implement at the level of the intermediate-representation of the compiler are shown in the following table: 
\newcommand{\pluseq}{\mathrel{+}=}

\begin{table}[H]
\caption{Compiler transformations that our approach uses to optimize the cost and constraint functions.}
\label{tab:transform}
\begin{tabular}{lll}
\toprule
 & \begin{minipage}{0.4\columnwidth}
\centering Original Code 
\end{minipage}
 & \begin{minipage}{0.4\columnwidth}
\centering Symbolic expression\end{minipage}\\
\midrule
1 & 
\begin{minipage}{0.4\columnwidth}
\begin{small}
\begin{algorithmic}
	\For {$i \gets 0$ to $N$}
		\State $v \pluseq const$
	\EndFor
\end{algorithmic}
\end{small}
\end{minipage}
& 
\begin{minipage}{0.4\columnwidth}
\begin{small}
\begin{algorithmic}
		\State $v \pluseq const \cdot N$
\end{algorithmic}
\end{small}
\end{minipage} 
\vspace{.1cm}\\
\midrule
2  & 
\begin{minipage}{0.4\columnwidth}
\begin{small}
\begin{algorithmic}
\For {$i \gets 0$ to $N$}
		\State $v \pluseq f(i)$
	\EndFor
\end{algorithmic}
\end{small}
\end{minipage} 
& 
\begin{minipage}{0.4\columnwidth}
\begin{small}
\begin{algorithmic}
		\State $ v \pluseq \sum_i{ f(i)}$
\end{algorithmic}
\end{small}
\end{minipage} \vspace{.3cm}\\
& 
\begin{minipage}{0.42\columnwidth}
\begin{small}
\begin{algorithmic}
\For {$i \gets 0$ to $N$}
		\State $v \pluseq \min(g(i), h(i))$
	\EndFor
\end{algorithmic}
\end{small}
\end{minipage} 
& 
\begin{minipage}{0.44\columnwidth}
\begin{small}
\begin{algorithmic}
		\State $ v \pluseq \sum_i{ \min(g(i), h(i))} $
		\State $\leq$ 
		\State $ v \pluseq \min( \sum_i{ g(i)}, \sum_i{ h(i) }) $
\end{algorithmic}
\end{small}
\end{minipage} \vspace{.3cm} \\
 & 
\begin{minipage}{0.4\columnwidth}
\begin{small}
\begin{algorithmic}
\For {$i \gets 0$ to $N$}
		\State $v \pluseq i^p$
	\EndFor
\end{algorithmic}
\end{small}
\end{minipage} 
& 
\begin{minipage}{0.44\columnwidth}
\begin{small}
\begin{algorithmic}
		\State $ v \pluseq \sum_i{ i^p}$ 
		\State \scriptsize{$(p+1)^{\rm th}$ degree polynomial}
		\State {derived from Faulhaber's formula}
\end{algorithmic}
\end{small}
\end{minipage} \vspace{.1cm} \\
\midrule
3 & 
\begin{minipage}{0.44\columnwidth}
\begin{small}
\begin{algorithmic}
		\If {$(\dots) $}{%
		\;$expr1$ }
		\Else {%
		\;$expr2$}
		\EndIf
\end{algorithmic}
\end{small}
\end{minipage} 
&
\begin{minipage}{0.44\columnwidth}
\begin{small}
\begin{algorithmic}
		\State $\max(expr1, expr2)$
\end{algorithmic}
\end{small}
\end{minipage}  \\
\bottomrule
\end{tabular}
\end{table}

\floatstyle{ruled}
\restylefloat{algorithm}

In particular, our optimization routine would:
\begin{enumerate}
\item Check if there is an addition between a variable $v$ initialized outside the loop and a loop invariant, and if $v$ is not used elsewhere in the loop. If so, apply transformation 1, where $N$ is the number of iterations of the loop.
\item Check if there is an addition between a variable $v$ initialized outside the loop and a function $f$ only depending on the inductive variable and other loop invariants. If so, apply transformation 2. In the particular case where $f(i)$ is $\min()$, the expression can be upper bounded as shown in Table~\ref{tab:transform}. In addition, polynomial expressions can be derived from some finite series using Faulhaber's formula~\cite{CG95}. 
\item Transform generic branching instructions into $\max(if, else)$ instructions, where the branch that gives the largest contribution to the cost function is selected.
\end{enumerate}

The following example shows the described transformations applied to the AQFT quantum algorithm. 
\begin{example}
The pseudo-code of the function $AQFT\_T$, which computes the total number of $T$ gates required for the AQFT algorithm, obtained after source-to-source transformation is shown in Alg.~\ref{qft_T}. As our implementation uses three rotation gates for each controlled-rotation, the function takes as input three accuracy parameters. This is why there are three innermost loops in Alg.~\ref{qft_T}. Since we want to extract a symbolic expression for the variable $T_{count}$, we have to get rid of as many loops as possible. 
The if statement in the inner-loop may be hoisted, resulting in the following expression:
\begin{table}[H]
\centering
\begin{minipage}{.4\textwidth}
\begin{small}
\begin{algorithmic}
  \For {$j \gets 0 $ to $\min(n-1-i, l)$}
\EndFor
\end{algorithmic}
\end{small}
\end{minipage}
\end{table}

Then all the loops are optimized by applying the transformations in Table~\ref{tab:transform}. Finally, the code in Alg.~\ref{qft_p} is obtained, which shows the closed-form expression for the $T_{count}$ with respect to the algorithm's parameters.
\end{example}

\noindent While our methodology succeeds at extracting closed-form expressions for all our examples, we note that this is not necessary for our methodology to work: the remaining control flow would not affect applicability or correctness, but merely cause an increase in runtime. Indeed, there are cases in which some residual code remains in the resulting expressions. This happens, for example, when some of the program parameters are read from a file. Consider the program performing phase estimation of the time evolution of a TFIM Hamiltonian described in Section~\ref{sec:sample}, but where the parameters $J$, $h$ and $n$ are read from a file. The program may check whether the input Hamiltonian is valid, e.g., whether $n\geq0$, before instantiating the circuit. As a consequence, such an if/else statement would remain in the final expression.

\begin{algorithm}[t]
\begin{algorithmic}
\Function{N\_ROT}{$\varepsilon_{rot}$}
	\Return $1.5 * \log_2(1./\varepsilon_{rot})$ 
\EndFunction
\Function{AQFT\_T}{$\varepsilon_{QFT}$, $\varepsilon_{R_1}$, $\varepsilon_{R_2}$, $\varepsilon_{R_3}$}
        \State $T_{count} \gets 0$
        \State $l \gets \lceil \log_2(n / \varepsilon_{QFT})\rceil +3$
        \For{$i\gets0$ to $n$ }
    		\For{$j \gets 0$ to $n-1-i$}
    			\If{$j\leq l$}
    			\For{$k \gets 0$ to $N\_ROT(\varepsilon_{R_1})$}
    			$T_{count}$++
    			\EndFor
    			\For{$k \gets 0$ to $N\_ROT(\varepsilon_{R_2})$}
    			$T_{count}$++
    			\EndFor
    			\For{$k \gets 0$ to $N\_ROT(\varepsilon_{R_3})$}
    			$T_{count}$++
    			\EndFor
    			\EndIf
    		\EndFor   		
   		\EndFor
        \Return $T_{count}$
\EndFunction  
\end{algorithmic}
\caption{Cost function for the AQFT algorithm}
\label{qft_T}
\end{algorithm}
\begin{algorithm}[t]
\begin{algorithmic}
\Function{AQFT\_T}{$\varepsilon_{QFT}$, $\varepsilon_{R_1}$, $\varepsilon_{R_2}$, $\varepsilon_{R_3}$}
       \State $T_{count} \gets 0$
        \State $l \gets \lceil \log_2(n / \varepsilon_{QFT})\rceil+3 $
    	 \State $T_{count} = T_{count} + \min( \frac{n(n-1)}{2} , nl) \cdot(N\_ROT(\varepsilon_{R_1})+N\_ROT(\varepsilon_{R_2}) + N\_ROT(\varepsilon_{R_3}))$
	\State\Return $T_{count}$ 
 \EndFunction
\end{algorithmic}
\caption{Cost function for the AQFT algorithm after loop optimization}
\label{qft_p}
\end{algorithm}

\section{Compiler and Language Requirements}
\label{sec:comp}

To add our methodology for automatic accuracy management to any quantum programming language, a few features must be added to the compiler if not already supported. Here we list all the features that the compiler must support to bring such an integration into fruition. 

We have identified such features by working on our initial prototype. The prototype has been developed using the LLVM project, which we chose for its modular infrastructure and libraries. The strategy that we identified while implementing our methodology in LLVM can be applied to any other quantum programming language. We illustrate this by developing a second prototype in Q\#. Furthermore, our LLVM prototype shows that support for our framework may also be added to quantum programming languages that are embedded in a classical host language.

\paragraph{{Don't Cares.}} Our methodology requires that the compiler identifies subroutine parameters that do not or only negligibly affect the total error and the cost function. We call such function parameters \textit{don't cares}. The corresponding arguments will be replaced by a default value in all calls to that subroutine. This allows the compiler to optimize repeated calls to the same function.
\begin{example} Consider AQFT, which includes many calls to the rotation gate with different rotation angles. Normally, these calls will have to be evaluated several times by the compiler, even if the angle will have no impact on the approximation error selected to decompose the rotation. We address this problem by annotating the angle parameter as a \textit{don't care}. This will result in many calls to the same function with identical arguments that hence may be removed by the compiler. 
\end{example}
\paragraph{{Epsilon Declarations.}} 
To provide language support (see Section~\ref{sect:support}) the compiler must be capable of locating all introduced accuracy parameters. This can be done by matching against a specific class or by using annotations.
\paragraph{{AST Transformation.}}\label{sect:astmod}
The compiler must provide access to the AST (or to a reasonably high-level IR) and allow rewriting and copying. In particular, we need to generate 3 different versions of the entire program: one that computes the total approximation error, one that computes the total cost, and the original quantum program, which will ultimately be invoked using optimized accuracy parameters.

The programs obtained after the described AST modifications could already be used to estimate the resource requirements of the quantum program. In this case, the only advantage with respect to using state-of-the-art methods would be that our approach provides language support by keeping track of all the introduced approximation errors, work that otherwise would have to be performed manually. In addition, the estimation would be too slow to be used in an optimization procedure (see results in Fig.~\ref{fig:plots}). As a remedy, we introduce specific rewrites that reduce the time to evaluate these expressions (see next paragraph).

\paragraph{{Rewrites to Make Evaluation More Efficient.}}\label{subs:rewrites} 
In order to speed up the optimization process, we need fast evaluations of the cost and error functions. The two functions extracted from most quantum algorithms will feature many loops performing simple counter increments or floating-point additions. 
To achieve our goal of extracting a symbolic expression from the control flow structure, we implement and employ several compiler optimizations, see Table~\ref{tab:transform}.

\subsection{LLVM Prototype}
Having described the features that are necessary to equip a programming language with automatic accuracy management, we now provide implementation details for our LLVM prototype.

\paragraph{Don't Cares}
In our LLVM implementation, we use compiler annotations to introduce additional information in the source code. In particular, we attach a \textit{don't care} annotation to a parameter declaration if it has negligible effect on the cost/constraint functions. 

\begin{figure}[t]
\begin{lstlisting}[style = cppstyle]
#include <qadf/epsilons.hpp>
#include <qadf/operations.hpp>

REGISTER_EPSILON(epsilon_QFT)

inline void Q_FUNC AQFT(int qubits[], const int n)
{
  epsilon_QFT eps_QFT;
  int lim = ceil(log2(n / eps_QFT))+3;
  for(int i = 0; i < n; i++)
  {
    H(qubits[i]);
    for (int j = 0; j < min(n-1-i, lim); j++)
    {
      CR(qubits[j + i + 1], M_PI/(1 << (j + 1)), qubits[i]);
    }
  }
}
\end{lstlisting}
	\caption{C++ code for the approximate QFT as consumed by our LLVM prototype.}
	\label{lst:aqftcpp}
\end{figure}
\paragraph{Epsilon Declarations}
Epsilon declarations are matched with a custom type. Our prototype framework provides a macro that allows the programmer to quickly define new types of error. This can be seen in the source code for the approximate QFT in Fig.~\ref{lst:aqftcpp}, where \texttt{epsilon\_QFT} is registered as a new type of error in line 4 and then used in lines 8--9.

\paragraph{AST Transformation}
We approach the problem of modifying the AST using a source-to-source transformation.
We implement a \textit{ClangTool} and run an \textit{ASTFrontendAction}: a routine that has access to the AST and allows us to interface with the source code. Our \textit{ClangTool} outputs files containing the function to compute the $T$-count, e.g., the one in Alg.~\ref{qft_T}, and the function computing the total approximation error $E$.
Our action exploits the \textit{ASTMatcher} library, which allows us to match nodes in the AST that have some specific properties. The library provides a concise way of describing patterns and is implemented as a \textit{domain-specific language} (DSL). 
In addition, matchers allow us to access the source code by running a callback function on the matched AST nodes. Once the locations of interest in the code have been identified, we can use an instance of the \textit{Rewriter} class to modify the code accordingly. As the AST is con\-stant by design, we must generate a new file containing our transformed source code.

Our tool also makes use of header files in which we define basic quantum operations, such as the ones in the Clif\-ford+${}T$ gate set. Those header files can be adjusted according to the specific application. For example, in addition to the $T$ gate, we might want to consider other expensive operations, e.g., two-qubit gates for NISQ devices.

The result of running the Clang tool is a new source file computing the total error or gate count as a function of all the accuracy parameters defined in the source code. 

\paragraph{Rewrites to Make Evaluation More Efficient}\label{loop-opt}
Our compiler optimization passes eliminate loops in the expressions for the cost (or the total error) by replacing them with additions and multiplications. For example, the first loop optimization described in Table~\ref{tab:transform} is supported for both integral and floating-point numbers. 
We use a custom LLVM \textit{loop pass} to perform this optimization. 
\begin{example}\label{excode}
Consider the following code that computes the total approximation error of $n$ quantum operations, characterized by the same approximation error $val$:

\begin{lstlisting}[style=cppstyle]
  double Eps = 0.00;
  double val = 0.02;
  for (int i = 0; i < n; i++) {
    Eps += val;
  }
\end{lstlisting}

Once this function is compiled into \textit{Intermediate Representation} (IR) code, $val$ will be identified as a loop-invariant variable, while $Eps$ will be assigned to a so-called PHI node. PHI nodes assign a variable with a different value, depending on the predecessor of the current block, where blocks are groups of instructions. In our example, the PHI node would have two incoming values: $0.00$  (if the predecessor block is outside the loop) and the temporary value containing the addition result  (if the predecessor was the previous loop iteration). PHI nodes are defined in the loop header block.
\end{example}
\noindent The loop pass traverses the code from the innermost to the outermost loop in the IR and checks whether:
\begin{enumerate}
\item it contains an instruction performing the addition operation between a loop-invariant operand and a variable defined through a PHI node in the loop header,
\item the PHI node is only used in the addition operation inside the loop or in the loop latch block,
\item the result of the addition is only used as the incoming value of the PHI node. 
\end{enumerate}
If the described conditions apply, the loop is removed. Referring to  the code in Example~\ref{excode}, the operations $\%1 = val * n$ and $\%2 = Eps + \%1$ would be added to the pre-header loop block, i.e., outside the loop. In addition, the result $\%2$ would replace all uses of the original addition operation, which can then be erased.
All the other transformations in Table 1 are implemented in a similar fashion.

\paragraph{Extraction of symbolic Representation.} Our LLVM prototype also has a method that navigates the fully optimized IR code and extracts symbolic expressions for the two estimates. The symbolic expression that we obtained by running our sample quantum algorithms are listed in Table~\ref{tab:expressions}.

The pass to extract the symbolic expression from the IR is implemented as an LLVM \textit{function pass}. Given the main function, it starts from the return instruction and recursively visits all instruction's operands annotating the respective functionality. The recursion terminates when the operand is a constant or, in general, not an instruction.

The extraction pass supports the following instructions: casting, PHI nodes, selects, truncations, zero extensions, call instructions, compare instructions, shifts, addition, multiplication, division, and subtraction. The expression is written in the Wolfram language, such that Mathematica~\cite{Mathematica} can be used for conversion to \LaTeX{} and further expression simplifications.

\subsection{Q\# Integration}
Q\# is a standalone quantum programming language developed by Microsoft to facilitates the description of hybrid quantum-classical programs. In Fig.~\ref{fig:qsharp-aqft}, we show a snippet of Q\# code that implements the approximate quantum Fourier transform operation, as described in Section~\ref{sec:sample}. 

\begin{figure}[t]
\begin{lstlisting}[style = qsharpstyle]
namespace Accuracy {

  open Microsoft.Quantum.Arrays;
  open Microsoft.Quantum.Convert;
  open Microsoft.Quantum.Intrinsic;
  open Microsoft.Quantum.Math;

  // intrinsic function for epsilon declaration
  function EpsilonValue() : Double {
    body intrinsic;
  }

  operation AQFT(qs : Qubit[]) : Unit is Adj+Ctl {
    let nQubits = Length(qs);
    let eps_QFT = EpsilonValue(); // epsilon-declaration
    let lim = Ceiling(Lg(IntAsDouble(nQubits) / eps_QFT)) + 3;
    for ((i, q) in Enumerated(qs)) { 
      H(q);
      for (j in 0 .. MinI(nQubits - i - 1, lim)) {
          (Controlled R1Frac)([q], (1, j + 1, qs[j + i + 1]));
      } 
    } 
  } 
}
\end{lstlisting}
\caption{Q\# implementation of the approximate quantum Fourier transform with epsilon declarations.}
\label{fig:qsharp-aqft}
\end{figure}

Qubits are represented in Q\# using the type \texttt{Qubit} and they are treated as opaque items that can be passed to both functions and operations, but that can only be interacted with by passing them to intrinsic (built-in) operations.
Q\# also uses namespaces to group definitions together, and elements from other namespaces may be referenced.
Q\# distinguishes functions from operations.  Functions are pure and free of side effects, whereas operations can have side effects, such as the application of an intrinsic operation to a qubit or register.
Q\# can perform type-safe symbolic computations to automatically derive the adjoint (inverse) and the controlled variants of an operation, enforced by providing the \texttt{is Adj+Ctl} declaration in line 13 of Fig.~\ref{fig:qsharp-aqft}. Line 15 shows a declaration of an approximation parameter, i.e., \texttt{eps\_QFT}, which is possible by adding language support for approximation errors. In the remainder of this section, we provide details on how this and other features have been implemented in the Q\# compiler.

\paragraph{Don't Cares}
Currently, our Q\# prototype does not support \textit{don't cares}. We implement
operations such as rotation gates---in which \textit{don't cares} can help to declare
that the rotation angle affects the gate cost only negligibly---as intrinsic
operations that introduce an epsilon variable and increment the gate counter
explicitly by a value depending on the accuracy parameter.  For more general cases,
we would require Q\# to support parameter-level annotations to declare
\textit{don't cares} explicitly.

\paragraph{Epsilon Declarations}
We use an intrinsic function for epsilon declarations.
Being intrinsic, it does not require an implementation inside the Q\# program to be used,
but we can locate it inside the AST in our transformation passes.
In the AQFT example, the function is declared in lines 9-11 and allows the programmer to declare and use accuracy parameters, e.g., as done in lines 15-16.

\paragraph{AST Transformation}
We have implemented an AST transformation pass to detect all \texttt{Epsilon\-Value} declarations, remove them, and add them as arguments to the operation signature.  This step is performed before producing the two pieces of code that compute the number of costly quantum gates and the upper bound on the overall approximation error.  Note that the intrinsic \texttt{EpsilonValue} function is never called in the resulting programs.

There exist no global variables and no call-by-reference parameters in Q\#.  However, it is possible to declare an intrinsic operation in Q\# and implement it in C\#.  In order to count the number of $T$ gates, we introduce an operation \texttt{Increment\-Counter(id, value)} whose implementation in C\# increments a global counter, called \texttt{id}, by \texttt{value}.

We use a similar technique as used for counting gates to accumulate the bound on the overall approximation error.  Each declaration of an epsilon value is replaced by a call to an operation \texttt{IncrementValue(id, value)} whose implementation in C\# increments a global variable called \texttt{id} by \texttt{value}.

\paragraph{Rewrites to Make Evaluation More Efficient}  
The Q\# compiler already contains some transformation passes, e.g., for operation inlining,
propagating constants, or removing unused code.  We have added two
additional transformation passes that optimize the use of
\texttt{IncrementCounter} and \texttt{IncrementValue} calls.  Without
loss of generality we explain the transformation passes by means of
the \texttt{Increment\-Counter} operation, remarking that they work
analogously for the \texttt{IncrementValue} operation.

The first transformation pass collects all \texttt{Increment\-Coun\-ter}
calls inside a scope level that have the same \texttt{id} and do not
contain values incorporating mutable variables.  These calls can be
merged into a single call by accumulating all values, benefiting from
further optimization, e.g., constant propagation.
The second transformation pass lifts \texttt{Increment\-Counter} calls
inside a for-loop.  If the call is the only statement in the body of
the for-loop and does not use the loop variable to compute the value,
the for-loop can be removed when multiplying the value in the
\texttt{IncrementCounter} call by the number of loop iterations.

\section{Qualitative Evaluation}\label{sect:qualitative}

In this section, we evaluate our prototypes using different quantum algorithms. Starting with a simple quantum Fourier transform, we increase the complexity of our examples. As a highlight, our prototype automatically extracts a symbolic expression for the phase estimation of a Trotter-decomposed time-evolution under a transverse-field Ising model Hamiltonian, where the phase estimation features an approximate QFT. The C++ code for the approximate QFT can be found in Fig.~\ref{lst:aqftcpp} and the corresponding Q\# code is shown in Fig.~\ref{fig:qsharp-aqft}. In addition, we also tested our prototype on Shor's algorithm~\cite{Shor94}, to provide the reader with a large-scale example. We used the period finding quantum routine as implemented in the open source programming framework ProjectQ~\cite{projectQ}.
All the obtained expressions for the $T$-count and the total approximation error are reported in Table~\ref{tab:expressions}.

\paragraph{Exact QFT}
The first example is an implementation of the exact quantum Fourier transform.
Our prototype is able to directly optimize all loops, including the outermost loop which yields a sum of the form $\sum_i^{n-1} i = \frac{n(n-1)}2$. It finds the correct closed-form expression for the total error and it correctly identifies the number of $T$ gates to be $\mathcal O(n^2\log(1/\varepsilon_R))$. For the detailed output, we refer to Table~\ref{tab:expressions}.

\paragraph{Approximate QFT (AQFT)}

Next, we consider the approximate quantum Fourier transform. The C++ source code that serves as the input to our prototype is depicted in Fig.~\ref{lst:aqftcpp}. Our prototype upper bounds an intermediate expression of the form $c+\sum_i{\min(f(i), g(i))}$ by choosing one of the arguments to the $\min$-function and successfully derives a closed-form expression for both the $T$ gate count and the total approximation error, see Table~\ref{tab:expressions}.

\paragraph{Quantum Phase Estimation (QPE)}\label{qpe_with_qft}
We combine the time evolution of a TFIM with QPE, to find the ground state of the TFIM.
In this first QPE example, we make the simplifying assumption that the inverse QFT can be performed natively. As can be seen in Table~\ref{tab:expressions} (labeled \textit{QPE simplified}), our methodology is capable of removing all loops and outputs two closed-form expressions for the $T$-count and the total error.

In a next step, we drop the simplifying assumption that the inverse QFT can be performed natively. We implement the inverse QFT as discussed in Section~\ref{sec:qft} and run our prototype on this larger example. The closed-form expressions for this case can be found in Table~\ref{tab:expressions} labeled \textit{QPE with QFT}.

Finally, we replace the exact QFT by an approximate QFT (see Fig.~\ref{lst:aqftcpp} for the C++ code). As in the AQFT example, our optimization pass upper bounds an intermediate expression of the form $c+\sum_i{\min(f(i), g(i))}$ by choosing one of the arguments to the $\min$-function. We refer to Table~\ref{tab:expressions} for the detailed output, which consists of two fully-symbolic expressions for the $T$-count and for the approximation error.

\DeclarePairedDelimiter\ceil{\lceil}{\rceil}
\afterpage{%
	\clearpage%
	\begin{sidewaystable}[!htbp]
		\Small
		\centering
		\caption{Symbolic expressions for the $T$-count (T) and total approximation error (E) as extracted by our prototype from the source code of various example programs.}\label{tab:expressions}
		        \begin{tabular}{ll}
        
        \toprule
            Algorithm (selected function) & Symbolic expression \\
            \midrule
            QFT  & $E = \frac{3}{2}\varepsilon_Rn(n-1) $ \\
             & $T \simeq 3.246n(n-1)\log\left(\frac{1}{\varepsilon_R}\right) $\\
            \midrule
            AQFT  & $E \simeq 3 \varepsilon_R n \left(\ceil*{\frac{\log(\frac{n}{\varepsilon_{QFT}})}{\log2}}+ 3\right)  + \varepsilon_{QFT}$  \\ 
              & $T\simeq 6.492 n \log\left( \frac{1}{\varepsilon_R} \right)
\left(\ceil*{\frac{\log(\frac{n}{\varepsilon_{QFT}})}{\log2}}+ 3\right) $\\
\midrule
			QPE simplified & $E \simeq \frac{ \left( \varepsilon_R (n-\frac{1}{2}) + \frac{\varepsilon_{TE}^{3/2}}{4}\right)
 \left( 2^{\ceil*{2.652 - 1.443 \log(\varepsilon_{QPE})} + 4}-4 \right)}{\sqrt{\varepsilon_{TE}}}+\varepsilon_{QPE}$\\
             & $T \simeq 
\frac{34.625 \left( n-\frac{1}{2} \right) 
\log \left( \frac{1}{\varepsilon_R} \right)
(2^{\ceil*{2.652 - 1.443 \log(\varepsilon_{QPE})} -\frac{1}{4}})}{\sqrt{\varepsilon_{TE}}}$ \\          
\midrule
            QPE with QFT & $E \simeq \frac{16\left( \varepsilon_R \left(n-\frac{1}{2} \right) +\frac{\varepsilon_{TE}^{3/2}}{4}\right) \left(2^{\ceil*{2.652-1.443 \log(\varepsilon_{QPE})} }-\frac{1}{4}\right)}{\sqrt{\varepsilon_{TE}}}+\frac{3}{2} \varepsilon_R (\ceil*{2.652-1.443 \log(\varepsilon_{QPE})} + 1)
 (\ceil*{2.652-1.443 \log(\varepsilon_{QPE})} +2) + \varepsilon_{QPE}$ \\
            & $T \simeq \log \left(\frac{1}{\varepsilon_R} \right) 
\Bigg(\frac{(n-\frac{1}{2}) ( 34.625^{\ceil*{2.652-1.443 \log( \varepsilon_{QPE} )}}-8.656 )}{\sqrt{\varepsilon_{TE}}} +
\ceil*{2.652-1.443 \log(\varepsilon_{QPE})}
(3.246 \ceil*{2.652-1.443 \log(\varepsilon_{QPE})} +
9.738) +6.492 \Bigg)$\\
\midrule
            QPE with AQFT  & 
            $E \simeq  3\varepsilon_R ( \ceil*{2.652-1.443 \log(\varepsilon_{QPE})} +2\left(\ceil*{\frac{\log(\frac{\lceil 2.652-1.443 \log(\varepsilon_{QPE}) \rceil+2}					{\varepsilon_{QFT}})}{\log2}}	 + 3 
\right) +\frac{16 \left(\varepsilon_R\left(n-\frac{1}{2}\right) + \frac{\varepsilon_{TE}^{3/2}}{4}\right)\left(2^{\ceil*{2.652-1.443 \log(\varepsilon_{QPE})}}-\frac{1}{4}\right)}{\sqrt{\varepsilon_{TE}}}  +
 \varepsilon_{QFT} + \varepsilon_{QPE}$ \\
            & 
            $T \simeq \log \left(\frac{1}{\varepsilon_R}\right) \Bigg(6.492 (\ceil*{2.652-1.443 \log(\varepsilon_{QPE})} + 2)\bigg( \ceil*{
\frac{
\log \left(\frac{\lceil 2.652-1.443 \log(\varepsilon_{QPE}) \rceil+2}{\varepsilon_{QFT}}\right)}{\log2}} + 3 \bigg) +\frac{34.625 \left(n-\frac{1}{2}\right) \left(2^{\ceil*{2.652-1.443 \log(\varepsilon_{QPE})}}-\frac{1}{4}\right)}{\sqrt{\varepsilon_{TE}}} \Bigg) $\\
\midrule
            Shor's algorithm~\cite{beauregard02} & 
            $E \simeq 4 n^2 \left( \varepsilon_R (n+1) \left( 30 \ceil*{ \frac{\log \left( \frac{n+1}{\varepsilon_{QFT}} \right)}{\log2}} + 109 \right) +10 \varepsilon_{QFT} \right)$ \\
            &   
            $ T \simeq 2n^2  \Bigg( 129.843(n+1) \log \left( \frac{1}{\varepsilon_R} \right) 
\ceil*{ \frac{\log\left(\frac{n+1}{\varepsilon_{QFT}}\right)}{\log2}} + 471.761(n+1)\log\left(\frac{1}{\varepsilon_R}\right) + 84n + 91 \Bigg)$\\
\bottomrule
        \end{tabular}
	\end{sidewaystable}
	\clearpage
}

\paragraph{Shor's Algorithm}
As last example, we present the results for Beauregard's implementation of Shor's algorithm~\cite{beauregard02}, which defines two approximation errors, one for the rotation gates $\varepsilon_R$ and one for the approximate QFT ($\varepsilon_{QFT}$). In this implementation, each controlled unitary in the phase-estimation procedure is a modular multiplication (by a constant). Please note that in our QPE example, we perform a phase-estimation on the time-evolution operator that evolves the system according to the transverse-field Ising model Hamiltonian. Therefore, while both examples make use of phase estimation, their cost and error functions are vastly different because the phase estimation is performed on different unitaries.
\\

\noindent In summary, our methodology successfully produces closed-form expressions for the total error and gate count for all our examples. In the next section, we show that access to symbolic expressions enables a significant reduction in the time required to optimize accuracy parameters.

\section{Quantitative Evaluation}\label{sect:runtime}

In this section, we demonstrate how our prototype implementation enables faster evaluations of the cost and constraint functions by leveraging the transformations in Table~\ref{tab:transform}. We compare the runtimes of our symbolic resource estimation method against a \textit{non-symbolic} approach. The latter does not use symbolic estimations for resources and errors. The non-symbolic estimates are generated by our prototypes during the AST modification step (see Section~\ref{sect:astmod}).

We ran all experiments on a MacBook Pro with an Intel Core i5 processor with 3.1 GHz processor clock frequency and 16 GB main memory. All source codes have been compiled using Clang version 9.0.0 with level 3 optimization (\emph{-O3}) and with the fast-math mode enabled (\emph{-ffast-math}).

\begin{figure*}[t]
\centering
\includegraphics[width=.7\textwidth]{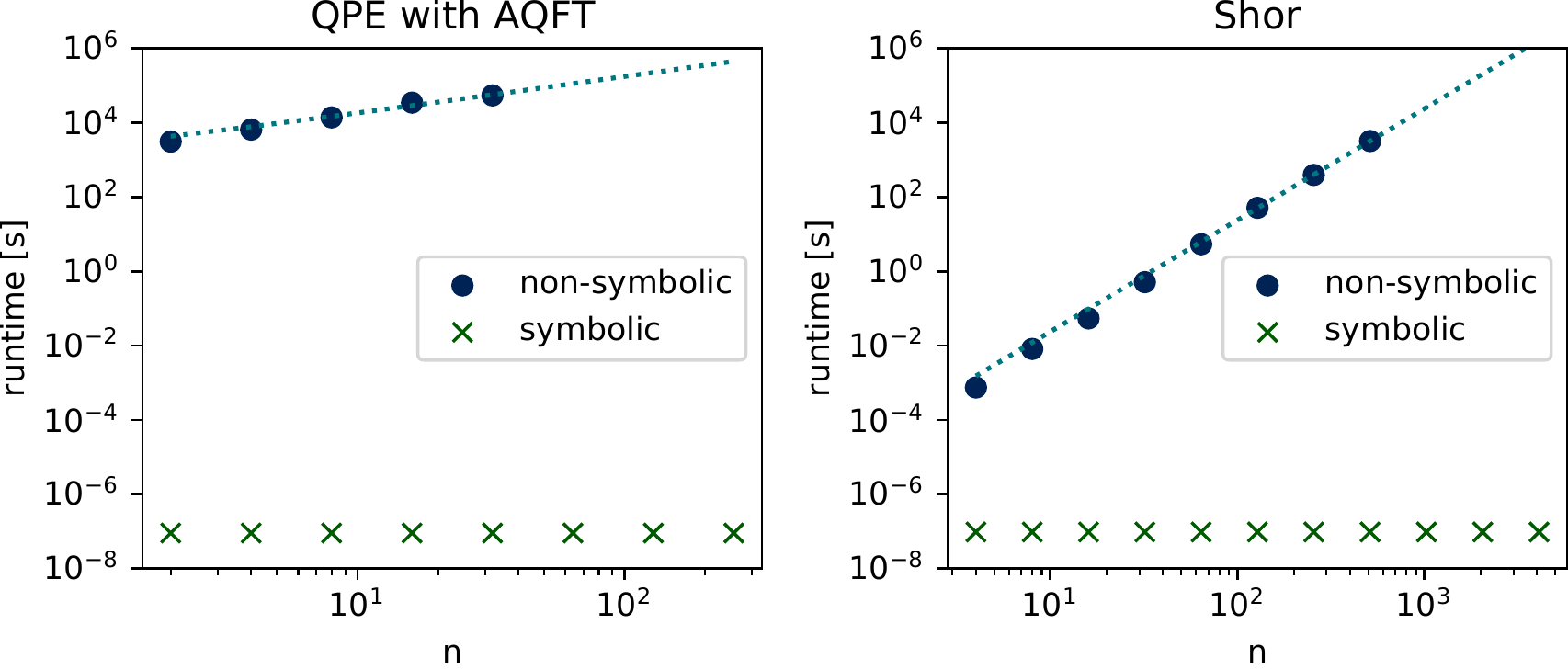}\hfill
\caption{Each data point marks the runtime required for a single evaluation of the $T$ function plus a single evaluation of the $E$ function for the QPE algorithm with approximate QFT and for Shor's algorithm. Optimized accuracy parameters are used as input in order to assert that the total approximation error is at most $5 \cdot 10^{-3}$. A comparison between the runtimes of the non-symbolic approach and the symbolic approach developed in this paper is shown.  We provide polynomial extrapolations from the collected runtimes for the non-symbolic approach as the runtime tops out for bit sizes above $512$ for Shor and $32$ for QPE.}
\label{fig:plots}
\end{figure*}

For these experiments, we choose our Clang/LLVM prototype. As previously described, it uses a two-mode annealing procedure to find suitable assignments for all accuracy parameters. 
We measure the runtime for performing one evaluation of the $T$ and $E$ functions using accuracy parameters provided by the annealer, guaranteeing an upper bound on the overall approximation equal to $5 \cdot 10^{-3}$. The runtime measurements are performed using different input sizes for Shor's algorithm and for our QPE example (with approximate QFT). The plots in Fig.~\ref{fig:plots} depict the sum of the runtimes required for evaluating the constraint and cost functions once, as they are always evaluated the same number of times in the simulated annealing procedure.
While the non-symbolic approach exhibits a growing runtime as a function of the problem size, our symbolic method shows the expected constant behavior. In particular, the runtime of the non-symbolic approach grows linearly for the QPE example and (roughly) cubically for our implementation of Shor's algorithm.

Given that both examples are valid applications of quantum computers only for large problem sizes (e.g., the target number of bits for Shor's algorithm is 
$n \approx 4000$), we show function extrapolations for the non-symbolic approach in Fig.~\ref{fig:plots}. The two resulting functions are $f_{QPE} = 1737.30x +816.98$ and $f_{Shor} = 2.38\cdot10^{-5} x^3$.

To estimate the time it would take to optimize accuracy parameters using a \textit{non-symbolic} approach, we multiply our runtime results in Fig.~\ref{fig:plots} by a lower bound on the number of function evaluations. For our examples, we find a loose lower bound of 100 evaluations (of each function). To determine suitable accuracy parameters for Shor's algorithm, we thus obtain a theoretical runtime of approximately $1890$ days for $n=4096$ bits and an overall approximation error of at most $5 \cdot 10^{-3}$.

Therefore, we may conclude that non-symbolic approaches are not suitable for large-scale applications.

\section{Conclusion and Outlook}
We describe the first framework with the ability to automatically manage approximation errors and outputting (near-)symbolic resource estimates.
Our methodology can be added to any quantum software framework, thereby greatly facilitating resource estimation of quantum programs. Such integration will allow even domain experts from, e.g., chemistry or machine learning, to write accuracy-aware quantum programs without having to manually derive and prove error bounds.

In this work, we identify the features that a quantum programming language must support in order to enable our methodology. We develop two prototype implementations that are capable of handling several example programs, including an implementation of Shor's algorithm.

Future work could implement improved handling of branching on measurement results and repeat-until-success-like structures. To handle such programs, our methodology requires additional input such as the maximal or expected iteration count (e.g., as a program annotation). For verification purposes, one could instrument the code in order to assert that the actual number of iterations does not deviate (too much) from the provided estimate.
Future work could also compare upper bounds to actually achieved errors on example applications. Currently, our methodology does not take into account gate cancellations that may be performed by circuit optimization. This, in addition to the repeated use of the triangle inequality to bound the overall error, likely leads to pessimistic error bounds. 

\begin{acks}
We thank the shepherd, Robert Rand, and the anonymous reviewers for their helpful suggestions and feedback, which significantly improved this paper. We are grateful to Vadym Kliuchnikov for enlightening discussions and comments.
\end{acks}


\end{document}